# Automated Tinnitus Detection Through Dual-Modality Neuroimaging: EEG Microstate Analysis and Resting-State fMRI Classification Using Deep Learning

Kiana Kiashemshaki[a], Sina Samieirad[b], Sarvenaz Erfani[c], Aryan Jalaeianbanayan[d], Nasibeh Asadi Isakan[e], Hossein Najafzadeh[f*]

[a] Master of Science in Computer Science, Department of Computer Science, Bowling Green State University, Bowling Green, Ohio, USA

[b] Department of Otolaryngology, Qaem Hospital, Mashhad University of Medical Sciences, Mashhad, Iran

[c] Department of Electrical Engineering, Amirkabir University of Technology, Tehran, Iran

[b] Department of Computer science, university of Verona, verona, Italy

[e] Department of Biomedical Engineering, University of Kentucky, Lexington, USA

[f] Department of Medical Bioengineering, Faculty of Advanced Medical Sciences, Tabriz University of Medical Sciences, Tabriz, Iran

**Authors:**

- Kiana Kiashemshaki
  Master of Science in Computer Science, Department of Computer Science, Bowling Green State University, Bowling Green, Ohio, USA
  E-mail: kkiana@bgsu.edu
  Orcid ID: https://orcid.org/0009-0001-5055-5552

- Sina Samieirad
  Department of Otolaryngology, Qaem Hospital, Mashhad University of Medical Sciences, Mashhad, Iran
  E-mail: sina.samieirad@gmail.com
  Orcid ID: https://orcid.org/0009-0005-6124-9162

- Sarvenaz Erfani
  Department of Electrical Engineering, Amirkabir University of Technology, Tehran, Iran
  E-mail: erfanisarvenaz@gmail.com

- Aryan Jalaeianbanayan
  Department of Computer science, university of Verona, verona, Italy
  E-mail: aryan.jalaeianbanayan@studenti.univr.it

- Nasibeh Asadi Isakan
  Department of Biomedical Engineering, University of Kentucky, Lexington, United States
  E-mail: nas237@uky.edu

- Hossein Najafzadeh
  Department of Medical Bioengineering, Faculty of Advanced Medical Sciences, Tabriz University of Medical Sciences, Tabriz, Iran
  E-mail: najafzadeh@tbzmed.ac.ir
  Orcid ID: https://orcid.org/ 0000-0003-2113-318x

**Corresponding Author:**

Dr. Hossein Najafzadeh

Department of Medical Bioengineering, Faculty of Advanced Medical Sciences, Tabriz University of Medical Sciences, Tabriz, Iran

E-mail: najafzadeh@tbzmed.ac.ir

Orcid ID: https://orcid.org/0000-0001-7411-8480

**Abstract**

**Objective:** Tinnitus affects 10-15% of the global population, yet lacks objective diagnostic biomarkers. This study evaluated machine learning approaches for tinnitus classification using electroencephalogram (EEG) and functional magnetic resonance imaging (fMRI) data to identify neural signatures distinguishing tinnitus patients from healthy controls.

**Methods:** Two independent datasets were analyzed: 64-channel EEG recordings from 80 participants (40 tinnitus patients, 40 controls) and resting-state fMRI data from 38 participants (19 tinnitus patients, 19 controls). EEG analysis employed comprehensive microstate feature extraction across four clustering configurations (4-state through 7-state) and five frequency bands (delta, theta, alpha, beta, gamma), yielding 440 features per participant. Additional analysis transformed Global Field Power signals into continuous wavelet transform images for deep learning classification. fMRI analysis utilized slice-wise convolutional neural network classification and hybrid architectures combining pre-trained models (VGG16, ResNet50) with traditional classifiers (Decision Tree, Random Forest, Support Vector Machine). Performance was evaluated using 5-fold cross-validation with accuracy, precision, recall, F1-score, and ROC-AUC metrics.

**Results:** EEG microstate analysis demonstrated systematic alterations in neural network dynamics among tinnitus patients, with the most pronounced changes observed in gamma-band microstate B occurrence rates (healthy: 56.56 vs tinnitus: 43.81 events/epoch, Cohen's d = 2.11, $p < 0.001$) and reduced alpha-band microstate coverage. Machine learning classification of microstate features yielded optimal performance with Random Forest and Decision Tree algorithms achieving 98.8% accuracy. Deep learning analysis of continuous wavelet transform-processed EEG signals revealed superior performance of VGG16 architecture across multiple frequency bands, with accuracies of 95.4% for delta and 94.1% for alpha frequencies. Resting-state fMRI analysis identified 12 high-performing axial slices (achieving ≥90% accuracy) from the total 32 evaluated slices, with slice 17 demonstrating optimal individual performance (99.0% ± 0.4% accuracy). The hybrid VGG16-Decision Tree model applied to combined high-performing slices achieved the highest classification accuracy of 98.95% ± 2.94% for automated tinnitus detection.

**Conclusion:** Both EEG and fMRI modalities provided effective neural signatures for tinnitus classification, with tree-based algorithms and hybrid architectures showing superior performance. The systematic neural alterations identified across multiple frequency bands and brain regions support the conceptualization of tinnitus as a multi-network disorder. Future research should focus on matched multimodal datasets to enable comprehensive feature fusion and enhance diagnostic accuracy.

# 1. Introduction

Tinnitus, commonly described as the perception of ringing, buzzing, or hissing sounds in the absence of an external auditory stimulus, affects millions globally and poses significant challenges to healthcare systems. Epidemiological studies suggest that approximately 10-15% of the global population experiences tinnitus, with 1-2% reporting severe and debilitating cases that greatly impact daily functioning and quality of life [1]. Individuals with chronic tinnitus often struggle with concentration, sleep disturbances, heightened stress, and increased risks of anxiety and depression [2]. Despite its prevalence, the exact pathophysiological mechanisms remain unclear, and the absence of objective biomarkers and standardized diagnostic tools complicates clinical assessment and treatment. This underscores the urgent need for innovative research and therapeutic strategies to address the growing burden of tinnitus.

The classification of electroencephalogram (EEG) microstate signals and functional magnetic resonance imaging (fMRI) data in resting-state tinnitus patients has become a prominent area of research, especially with the increasing application of machine learning and deep learning techniques. These advanced methods have played a crucial role in uncovering the neural mechanisms underlying tinnitus, a condition marked by the perception of phantom auditory sensations in the absence of external stimuli [3].

EEG microstates represent transient patterns of brain activity that can be indicative of underlying cognitive processes. Recent research has emphasized the utility of microstate analysis as a valuable tool for identifying and classifying neurological disorders, such as tinnitus. For instance, Manabe et al. demonstrated that a microstate-based regularized common spatial pattern (CSP) approach achieved classification accuracies exceeding 90% in surgical training scenarios, suggesting that similar methodologies could be applied to classify EEG signals in tinnitus patients [4]. Furthermore, Kim et al. explored the use of EEG microstate features for classifying schizophrenia, indicating that machine learning could enhance the robustness of microstate analysis when combined with other neuroimaging modalities [5]. This underscores the versatility of microstate features across different neurological contexts.

Recent developments in deep learning have yielded significant progress in various medical fields, notably in the diagnosis of neuropsychiatric disorders and biomedical classification tasks. Additionally, integrated models that leverage transformers, generative adversarial networks (GANs), and traditional neural architectures have proven highly effective in addressing complex problems within the information technology sector [6-11]. In tinnitus research, Hong et al. demonstrated that deep learning models like EEGNet outperform traditional SVMs by automatically learning complex EEG signal patterns, highlighting their potential for improved diagnostics [12]. Saeidi et al. reviewed neural decoding with machine learning, emphasizing the superior classification performance of deep learning across various EEG applications [13]. Temporal dynamics in EEG microstate analysis have also been explored, as Agrawal et al. used attention-enhanced LSTM networks to classify temporal cortical interactions, achieving impressive results with high classification accuracy [14]. Similarly, Keihani et al. showed that Bayesian optimization enhances resting-state EEG microstate classification in schizophrenia, suggesting its applicability to tinnitus research [15]. Furthermore, Piarulli et al. combined EEG data from 129 tinnitus patients and 142 controls to identify biomarkers using linear SVM classifiers. They achieved accuracies of 96% and 94% for distinguishing tinnitus patients from controls and 89% and 84% for differentiating high and low distress levels, with minimal feature overlap indicating distinct neuronal mechanisms for distress and tinnitus symptoms [16].

Recent studies have utilized fMRI data and advanced ML/DL techniques to improve tinnitus classification. Ma et al. highlighted the potential of fMRI in treatment evaluation, demonstrating that neurofeedback alters neurological activity patterns, although specific metrics were not provided [17]. Rashid et al. reviewed ML and DL applications in fMRI, emphasizing CNNs' ability to identify brain states associated with disorders like tinnitus [18]. Cao et al. achieved 96.86% accuracy in Alzheimer's classification using a hybrid 3D CNN and GRU network, showcasing adaptable methodologies for tinnitus research [19]. Lin et al. developed a multi-task deep learning model combining multimodal structural MRI to classify tinnitus and predict severity, successfully distinguishing tinnitus patients from controls [20]. Xu et al. applied rs-fMRI with CNNs to differentiate 100 tinnitus patients from 100 healthy controls, achieving an AUC of 0.944 on the Dos_160 atlas and highlighting functional connectivity's diagnostic value [21]. Pre-trained models like VGG16 and ResNet50 have further enhanced classification accuracy in fMRI data through transfer learning, addressing limitations in tinnitus datasets [22].

Previous studies on tinnitus diagnosis have primarily relied on either EEG or fMRI data independently, with limited exploration of comprehensive feature extraction and classification approaches. EEG studies often lack detailed microstate analysis across multiple clustering configurations (4 to 7 states) and comprehensive frequency band decomposition, while fMRI studies have not systematically investigated slice-wise analysis or hybrid model architectures combining pre-trained networks with traditional classifiers. These methodological gaps have restricted the development of robust and objective diagnostic methods for tinnitus assessment. The present study addresses these limitations through a dual-modality approach that independently analyzes EEG and fMRI data using advanced machine learning techniques. For EEG analysis, we implement comprehensive microstate feature extraction across four clustering configurations (4-state through 7-state) and five frequency bands (delta, theta, alpha, beta, and gamma), combined with continuous wavelet transform imaging for deep learning classification. For fMRI analysis, we employ slice-wise CNN classification and hybrid architectures that integrate pre-trained models (VGG16 and ResNet50) with traditional machine learning classifiers (Decision Tree, Random Forest, and SVM). While the datasets were derived from different participant cohorts preventing direct multimodal fusion, this independent dual-modality analysis provides complementary neuroimaging perspectives on tinnitus pathophysiology and establishes a methodological framework for future integrated multimodal investigations. This approach represents an advancement in objective tinnitus classification by demonstrating the effectiveness of comprehensive feature extraction and hybrid model architectures across both temporal and spatial neuroimaging domains.

## 2. Methods

### 2.1. Study Design and Dataset Overview

Neural alterations in tinnitus were investigated using two distinct neuroimaging modalities that were analyzed independently. Separate analyses of electrophysiological (EEG) and hemodynamic (fMRI) brain activity were conducted to provide complementary perspectives on tinnitus pathophysiology. Two independent datasets were examined: one EEG cohort for electrophysiological investigation and one fMRI cohort for hemodynamic network analysis.

### 2.2. Participants and Data Acquisition

#### 2.2.1. EEG Dataset

##### 2.2.1.1. Participant Characteristics and Clinical Assessment

The primary EEG cohort included two distinct groups: individuals with chronic tinnitus and neurotypical controls. The tinnitus group consisted of 40 participants (24 females, 16 males; mean age = 42.8 ± 13.2 years, range 24-68 years) experiencing persistent tinnitus symptoms for 1–30 years. The control group comprised 40 age-matched healthy volunteers (22 females, 18 males; mean age = 41.3 ± 12.8 years, range 25-66 years) with no tinnitus history or significant medical conditions. Groups were carefully age-matched to eliminate potential confounding effects of age-related neural changes on electrophysiological measures. Recruitment occurred through otolaryngology clinics and audiology centers following comprehensive clinical evaluations. All participants discontinued medications two weeks before testing and completed thorough medical screenings to confirm eligibility. Exclusion criteria included neurological conditions, active ear infections, or substantial hearing loss unrelated to tinnitus pathophysiology. Identical screening procedures were applied across groups to ensure methodological rigor and valid between-group comparisons.

Table 1 summarizes participant demographics, audiological parameters, tinnitus characteristics (laterality, frequency, subjective loudness), hearing thresholds, and standardized auditory indices for the tinnitus cohort. Psychological assessment encompassed validated measures of anxiety, depression, and tinnitus-related distress.

Table 1: Clinical and Audiological Characteristics of Primary Tinnitus Cohort

| ID | Sex | Age | Tinnitus Phenotype | Frequency (Hz) | Loudness (dB SL) | Hearing Threshold (dB HL) | Audiometric Index | Anxiety Score | Depression Score | Severity Rating | Laterality |
|---|---|---|---|---|---|---|---|---|---|---|---|
| **T01** | F | 52 | Tonal | 6100 | 32 | 12 | 17 | 2 | 2 | Mild | Left |
| **T02** | M | 35 | Tonal | 1550 | 14 | 7 | 9 | 2 | 1 | Minimal | Right |
| **T03** | M | 61 | Tonal | 10100 | 52 | 10 | 19 | 1 | 2 | Minimal | Left |
| **T04** | F | 45 | Complex | 8100 | 48 | 23 | 27 | 2 | 2 | Moderate | Left |
| **T05** | F | 29 | Tonal | 7900 | 16 | 17 | 22 | 1 | 1 | Minimal | Right |
| **T06** | F | 68 | Tonal | 11300 | 88 | 9 | 13 | 2 | 2 | Severe | Left |
| **T07** | M | 42 | Tonal | 1950 | 37 | 19 | 15 | 1 | 1 | Minimal | Right |
| **T08** | F | 38 | Complex | 7450 | 26 | 19 | 23 | 2 | 2 | Minimal | Left |
| **T09** | F | 31 | Tonal | 5900 | 11 | 14 | 15 | 1 | 1 | Severe | Right |
| **T10** | F | 56 | Tonal | 9100 | 42 | 18 | 25 | 2 | 2 | Severe | Left |
| **T11** | M | 49 | Tonal | 4050 | 12 | 19 | 20 | 3 | 2 | Minimal | Right |

| ID | Sex | Age | Phenotype | Frequency | Loudness | | | | | Severity | Side |
|---|---|---|---|---|---|---|---|---|---|---|---|
| T12 | M | 33 | Complex | 2950 | 24 | 14 | 13 | 1 | 1 | Absent | Left |
| T13 | F | 47 | Tonal | 4950 | 46 | 21 | 24 | 2 | 1 | Minimal | Right |
| T14 | M | 36 | Tonal | 2525 | 32 | 13 | 11 | 2 | 1 | Minimal | Right |
| T15 | F | 54 | Complex | 6950 | 49 | 15 | 19 | 1 | 1 | Severe | Left |
| T16 | F | 41 | Tonal | 10050 | 57 | 22 | 31 | 2 | 2 | Moderate | Right |
| T17 | M | 28 | Tonal | 2050 | 21 | 20 | 22 | 1 | 1 | Minimal | Left |
| T18 | F | 58 | Tonal | 6050 | 36 | 17 | 21 | 2 | 1 | Severe | Right |
| T19 | M | 39 | Tonal | 4050 | 16 | 13 | 14 | 1 | 1 | Minimal | Left |
| T20 | F | 44 | Complex | 8050 | 22 | 42 | 47 | 2 | 2 | Severe | Right |
| T21 | M | 32 | Tonal | 3200 | 28 | 15 | 18 | 2 | 1 | Mild | Left |
| T22 | F | 48 | Complex | 9200 | 35 | 18 | 24 | 1 | 2 | Moderate | Right |
| T23 | M | 26 | Tonal | 4800 | 19 | 11 | 14 | 1 | 1 | Minimal | Left |
| T24 | F | 53 | Tonal | 7200 | 44 | 21 | 26 | 2 | 2 | Severe | Right |
| T25 | M | 37 | Complex | 5400 | 31 | 16 | 20 | 2 | 1 | Mild | Left |
| T26 | F | 46 | Tonal | 8800 | 39 | 19 | 23 | 1 | 2 | Moderate | Right |
| T27 | M | 30 | Tonal | 2800 | 22 | 13 | 16 | 1 | 1 | Minimal | Left |
| T28 | F | 59 | Complex | 6600 | 47 | 24 | 29 | 2 | 2 | Severe | Right |
| T29 | M | 43 | Tonal | 3600 | 26 | 17 | 21 | 2 | 1 | Mild | Left |
| T30 | F | 34 | Tonal | 9600 | 33 | 14 | 18 | 1 | 1 | Moderate | Right |
| T31 | M | 51 | Complex | 4200 | 41 | 20 | 25 | 2 | 2 | Severe | Left |
| T32 | F | 40 | Tonal | 7800 | 29 | 16 | 22 | 1 | 1 | Mild | Right |
| T33 | M | 27 | Tonal | 5200 | 24 | 12 | 15 | 1 | 1 | Minimal | Left |
| T34 | F | 55 | Complex | 8400 | 45 | 22 | 27 | 2 | 2 | Severe | Right |
| T35 | M | 38 | Tonal | 3800 | 30 | 18 | 23 | 2 | 1 | Moderate | Left |
| T36 | F | 47 | Tonal | 6400 | 37 | 20 | 25 | 1 | 2 | Moderate | Right |
| T37 | M | 24 | Complex | 4600 | 25 | 14 | 17 | 1 | 1 | Minimal | Left |
| T38 | F | 60 | Tonal | 7600 | 43 | 25 | 30 | 2 | 2 | Severe | Right |
| T39 | M | 45 | Tonal | 5000 | 34 | 19 | 24 | 2 | 1 | Moderate | Left |
| T40 | F | 41 | Complex | 9000 | 40 | 17 | 21 | 1 | 2 | Moderate | Right |

*Note: Tinnitus phenotypes classified as "Tonal" (narrow-band, pure-tone characteristics) or "Complex" (broadband, multi-component features). Loudness expressed in dB sensation level (SL) above individual hearing thresholds. Severity ratings based on standardized clinical assessment scales.*

## 2.2.1.2. EEG Data Acquisition Protocol

EEG recordings were obtained using a 64-channel high-density acquisition system from both tinnitus patients and controls. Participants sat comfortably with eyes closed to minimize sensory confounds. Electrode positioning followed the international 10-10 standard, with Cz serving as the reference electrode and left earlobe as ground. All sessions occurred in an electromagnetically shielded, acoustically isolated chamber to optimize signal quality. Recording parameters included 1200 Hz sampling frequency with electrode impedances maintained below 50 kΩ. The electrode array connected to a GAMMABox amplification system. Each participant completed a 5-minute resting-state session to capture baseline neural oscillations consistently across experimental groups.

### 2.2.2. Functional MRI Dataset

### 2.2.2.1. Acoustic Trauma-Induced Tinnitus Cohort

A publicly accessible rs-fMRI dataset investigating acoustic trauma-induced tinnitus was incorporated to provide complementary neuroimaging perspectives. This cohort included 38 male participants: 19 with chronic tinnitus from acoustic trauma (mean duration: 11.9 ± 9.6 years; median: 12 years) and 19 age-matched healthy controls (mean age: 42.5 ± 11.9 years). Tinnitus participants were recruited through military medical centers and hospital otolaryngology departments, meeting strict criteria of persistent symptoms for ≥6 months post-acoustic exposure. Controls were recruited via community volunteers with comprehensive screening excluding tinnitus history or significant auditory pathology. Audiological evaluation employed standardized pure-tone audiometry on the MRI acquisition day. Hearing thresholds were measured at six frequencies (0.25, 0.5, 1, 2, 4, and 8 kHz) using calibrated Audioscan FX equipment conforming to AFNOR standards. Threshold determination followed the Hughson-Westlake ascending method, reported in dB HL. Tinnitus severity assessment utilized the validated Tinnitus Handicap Inventory (THI) providing standardized impact classifications. Table 2 presents comprehensive demographic, audiological, and clinical characteristics including age distribution, laterality, symptom duration, THI scores, trauma etiology, and bilateral hearing profiles [23].

Table 2: Demographic and Clinical Characteristics of fMRI Tinnitus Participants

| Subject | Age (years) | Laterality | Duration (years) | THI Score | Trauma Etiology | Right Ear Loss | Left Ear Loss |
|---|---|---|---|---|---|---|---|
| P1 | 55 | Bilateral | 11 | 7 | Occupational | 2-8 kHz | 0.25, 8 kHz |
| P2 | 44 | Bilateral | 26 | 9 | Military | - | 0.5 kHz |
| P3 | 39 | Bilateral | 14 | 24 | Military | 4-8 kHz | 4-8 kHz |
| P4 | 60 | Bilateral | 26 | 23 | Military | 4-8 kHz | 4-8 kHz |
| P5 | 29 | Bilateral | 11 | 9 | Musical | - | - |
| P6 | 41 | Bilateral | 3 | 15 | Military | 0.5-1, 8 kHz | 8 kHz |
| P7 | 42 | Bilateral | 15 | 31 | Military | 8 kHz | 8 kHz |
| P8 | 43 | Bilateral | 17 | 23 | Combined | 0.25-4 kHz | 2-8 kHz |
| P9 | 50 | Left | 11 | 45 | Military | 0.25-8 kHz | 0.25-8 kHz |
| P10 | 41 | Right | 17 | 17 | Military | - | - |
| P11 | 27 | Left | 2 | 13 | Military | - | 4-8 kHz |
| P12 | 59 | Bilateral | 18 | 11 | Military | 0.25-0.5 kHz | 0.25-8 kHz |
| P13 | 23 | Bilateral | 2 | 25 | Musical | - | - |
| P14 | 26 | Bilateral | 17 | 5 | Musical | - | - |
| P15 | 41 | Right | 3 | 13 | Musical | 4-8 kHz | 0.5, 8 kHz |
| P16 | 50 | Right | 0.5 | 13 | Military | 4-8 kHz | 8 kHz |
| P17 | 58 | Bilateral | 11 | 7 | Occupational | 4-8 kHz | 4-8 kHz |
| P18 | 50 | Left | 11 | 15 | Musical | - | - |
| P19 | 34 | Bilateral | 14 | 5 | Military | - | 0.25-0.5 kHz |

*THI: Tinnitus Handicap Inventory. Hearing loss frequencies indicate ranges with elevated thresholds (>20 dB HL).*

### 2.2.2.2. MRI Data Acquisition Protocol

Neuroimaging was performed using a 3.0 Tesla Philips Achieva-TX scanner with a 32-channel phased-array head coil optimized for brain imaging. The imaging protocol included extended resting-state functional sequences and high-resolution T1-weighted structural scans for anatomical reference and spatial normalization.

To enhance participant comfort and minimize scanner noise effects on tinnitus perception, specialized gradient echo EPI sequences with SoftTone noise reduction were implemented. This approach substantially reduced acoustic interference while preserving optimal image quality.

Functional acquisition parameters were: 32 axial slices, 3.5 mm thickness for whole-brain coverage, 3 × 3 mm² in-plane resolution, TR = 2000 ms, TE = 32 ms optimized for BOLD sensitivity, flip angle = 75°. A total of 400 volumes were acquired over 13 minutes 20 seconds, providing extensive temporal data for robust connectivity analyses. These parameters ensured superior signal-to-noise ratios while minimizing motion artifacts and physiological noise that could compromise connectivity measurements.

Table 3 summarizes the structural organization of the rs-fMRI dataset used in this study.

Table 3: rs-fMRI Dataset Structure and Organization

| Parameter | Value | Description |
| --- | --- | --- |
| **Total Subjects** | 38 | 19 tinnitus patients + 19 healthy controls |
| **Tinnitus Subjects** | 19 | Chronic tinnitus from acoustic trauma |
| **Control Subjects** | 19 | Age-matched healthy controls |
| **Axial Slices per Subject** | 32 | Whole-brain coverage with 3.5 mm thickness |
| **Time Points per Subject** | 400 | Acquired over 13 minutes 20 seconds |
| **Total Images per Subject** | 12,800 | 32 slices × 400 time points |
| **Total Dataset Images** | 486,400 | 38 subjects × 12,800 images |
| **Spatial Resolution** | 3 × 3 mm² | In-plane resolution |
| **Temporal Resolution (TR)** | 2000 ms | Repetition time |
| **Acquisition Duration** | 13:20 min | Total scanning time per subject |

## 2.3. Ethical Considerations

This study was performed in full compliance with the Declaration of Helsinki and all applicable ethical guidelines and standards. Ethical approval was granted by the Local Ethics Committee CPP Sud-est V (Reference: 10-CRSS-05 MS 14-52) for the rs-fMRI data, as well as by the Ethics Committee under trial registration number ISRCTN14553350 for the EEG dataset. Written consent was secured from all participants or their legal representatives, ensuring adherence to the inclusion criteria specified for both datasets.

## 2.4. Data Preprocessing

### 2.4.1. Preprocessing Pipeline Overview

To ensure the quality and consistency of the data used in this study, various preprocessing methods were applied to both EEG signals and rs-fMRI images. Each preprocessing technique was designed to address specific data challenges and enhance the analysis. Table 4 presents a comprehensive summary of the preprocessing techniques applied to both EEG and rs-fMRI datasets.

Table 4: Summary of Preprocessing Steps

| Modality | Preprocessing Step | Description | Equation |
|---|---|---|---|
| EEG | Data Segmentation and Filtering [24] | Segmented into 10-second epochs (1200 samples per epoch) and filtered using a 3rd-order Butterworth bandpass filter (0.5–50 Hz). | $H(s) = \dfrac{1}{\sqrt{1 + (\frac{s}{w_c})^{2n}}}$ |
| EEG | Normalization and DC Correction [25] | Normalized to $\mu$ = mean $(x)$, std $(\sigma)$= 1, with DC offset correction by subtracting the mean from each segment. | $x_{norm} = \dfrac{x - \mu}{\sigma}$ |
| EEG | Outlier Removal | Segments with voltage outside -150 to 150 µV were excluded to remove artifacts like eye blinks and muscle activity. | --- |
| EEG | Frequency Decomposition [24] | Decomposition of the EEG signal into Delta (0.5-4 Hz), Theta (5-8 Hz), Alpha (9-14 Hz), Beta (15-30 Hz), and Gamma (31-50 Hz) bands using the wavelet transform with the Daubechies 4 (Db4) wavelet. | $X(t) = \sum_{j=-\infty}^{\infty} \sum_{k=-\infty}^{\infty} c_{j,k} \psi_{j,k}(t)$ |
| rs-fMRI | Normalization | Scaled pixel intensity values to [0, 1] by dividing by 255. | $I_{norm} = \dfrac{I}{255}$ |
| rs-fMRI | Noise Reduction [26] | Applied a median filter (kernel size 3×3) to reduce noise while preserving edges. | $I_{\text{filtered}}(i,j) = median\ \{I(k,l)|(k,l) \in kernel\}$ |
| rs-fMRI | Contrast Enhancement [27] | Enhanced contrast using CLAHE with a 3×3 kernel to improve local contrast while preventing noise amplification. | $I_{\text{CLAHE}}(i,j) = \text{CLAHE}(I(i,j), kernel\ size = 3 \times 3)$ |

### 2.4.2. EEG Signal Preprocessing

EEG signals underwent comprehensive preprocessing including segmentation into 10-second epochs, bandpass filtering (0.5-50 Hz), normalization, outlier removal, and frequency decomposition into five bands (Delta, Theta, Alpha, Beta, Gamma) using Daubechies 4 wavelet transform. Figure 1 illustrates a sample of filtered EEG signals for all channels in one-second intervals, comparing healthy and tinnitus groups across different frequency bands.

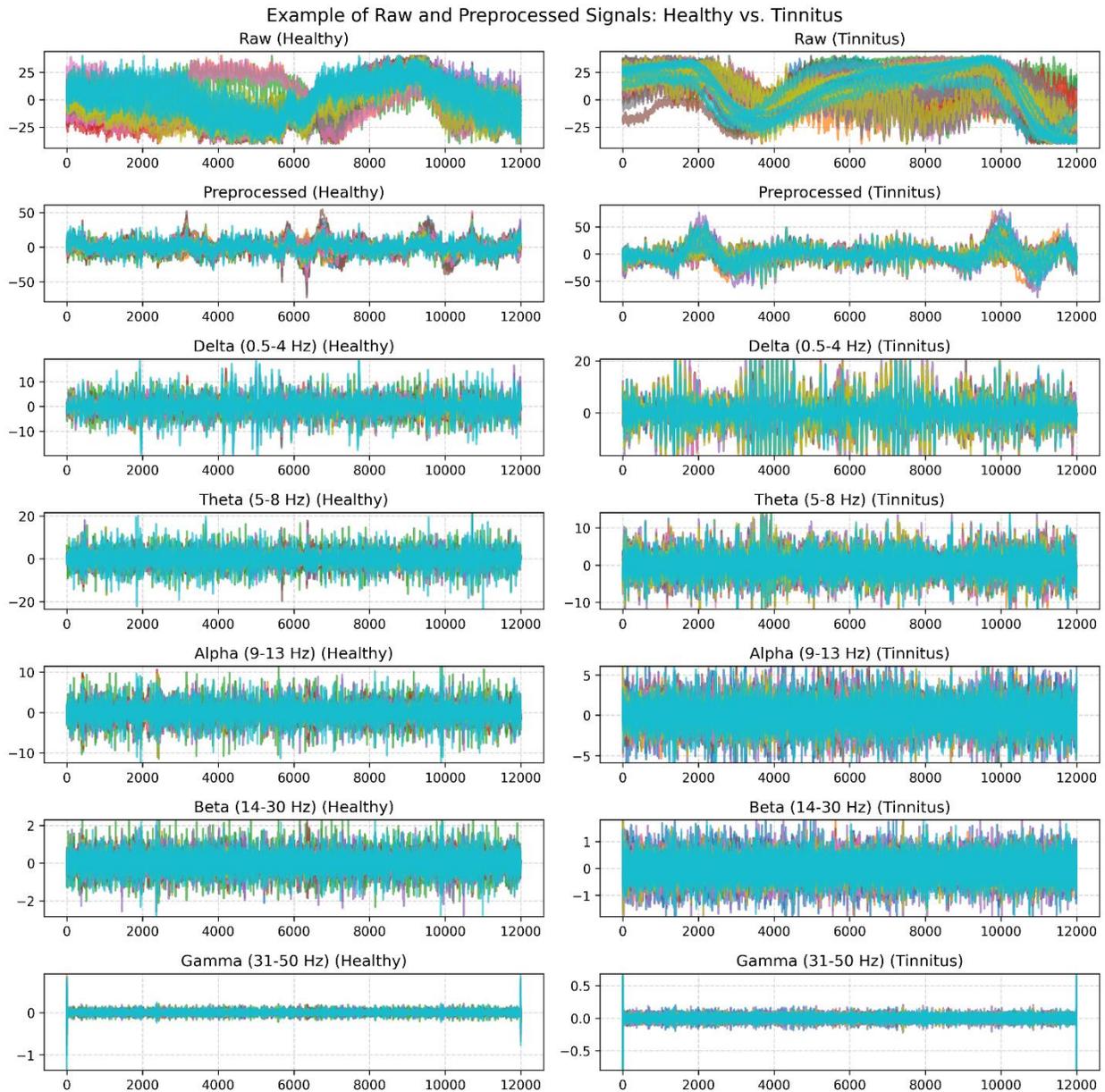

Figure 1: An Example of Filtered EEG Signals for One Second: Original and Band-Pass Filtered Signals in Delta, Theta, Alpha, Beta, and Gamma Bands Across Healthy and Tinnitus Groups.

### 2.4.3. fMRI Image Preprocessing

Rs-fMRI images were preprocessed through normalization, noise reduction using median filtering, and contrast enhancement using Contrast Limited Adaptive Histogram Equalization (CLAHE) to improve image quality and standardize intensity values. Figure 2 illustrates an example of improved rs-fMRI images of a patient after averaging 400 time points across 32 slices.

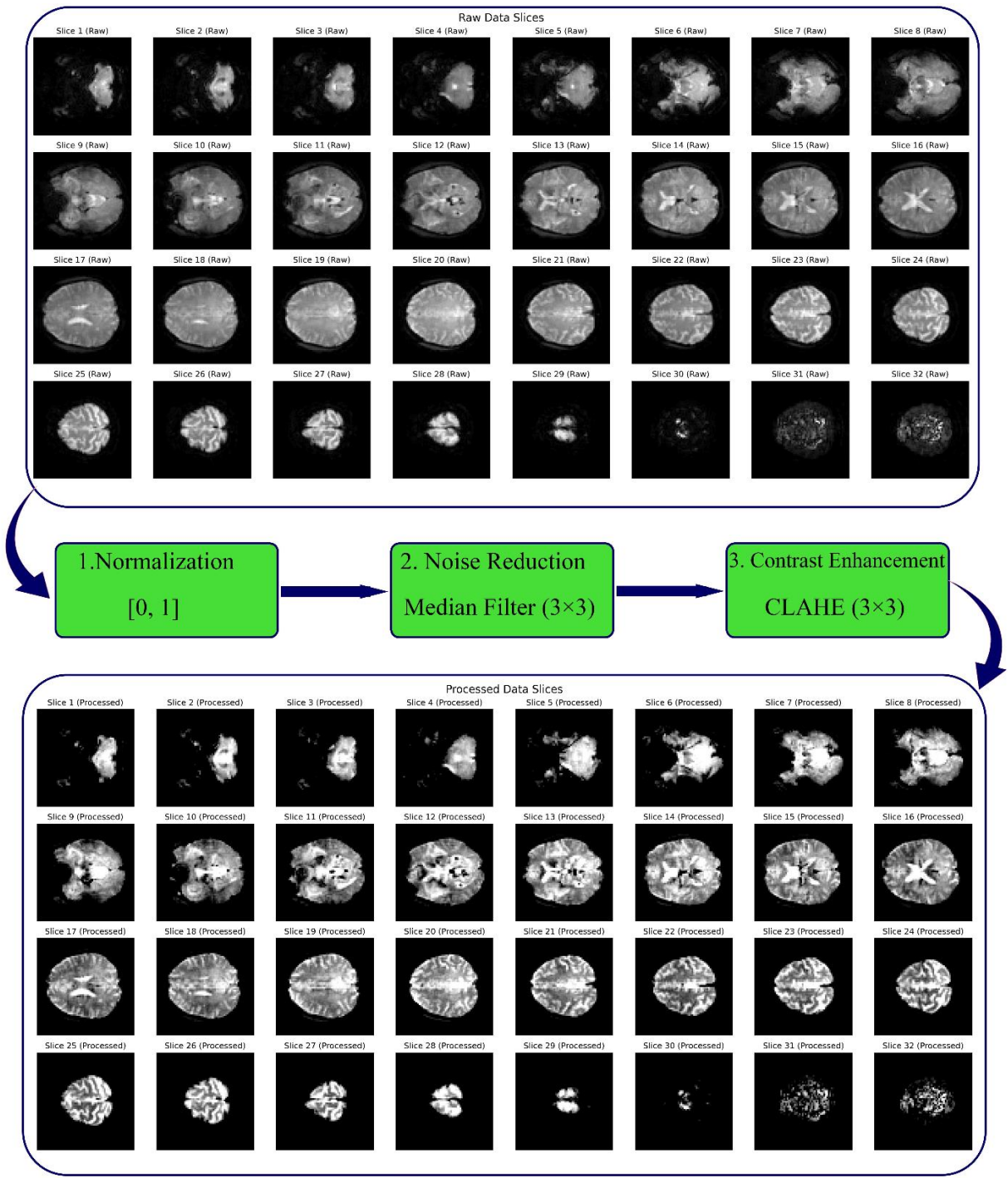

**Figure 2:** Comparison of Raw and Improved rs-fMRI Images After Image Enhancement Operations.

## 2.5. Feature Extraction

### 2.5.1. EEG Feature Extraction Methods

## 2.5.1.1. Microstate-Based Statistical Features

Features were extracted based on microstates corresponding to 4-state (A, B, C, D), 5-state (A, B, C, D, E), 6-state (A, B, C, D, E, F), and 7-state (A, B, C, D, E, F, G) models. The dataset comprised 80 participants (40 tinnitus patients and 40 healthy controls), with each participant's 5-minute resting-state EEG recording segmented into 10-second non-overlapping windows, yielding 300 windows per participant (total: 24,000 windows). The specifications are summarized in Table 5.

Table 5: Feature Extraction and Dataset Specifications

| Parameter | Specification | Details |
| --- | --- | --- |
| Participants | 80 | 40 tinnitus patients, 40 healthy controls |
| Recording Duration | 5 minutes | Resting-state EEG |
| Window Segmentation | 10-second non-overlapping | 300 windows per participant |
| Total Windows | 24,000 | 300 windows × 80 participants |
| Frequency Sub-bands | 5 bands | Delta (1–4 Hz), Theta (4–8 Hz), Alpha (8–13 Hz), Beta (13–30 Hz), Gamma (30–45 Hz) |
| Microstate Configurations | 4 models | 4-state (A–D), 5-state (A–E), 6-state (A–F), 7-state (A–G) |
| Microstate Parameters | 4 features per state | Duration, Coverage, Occurrence, Mean GFP |
| Features per Configuration | Varies by model | 4-state: 80, 5-state: 100, 6-state: 120, 7-state: 140 |
| Total Features per Window | 440 | Combined from all configurations |
| Final Dataset Dimensions | 24,000 × 440 | Windows × Features matrix |
| Classification Samples | 24,000 | 12,000 tinnitus, 12,000 control windows |

These microstates were identified using a modified K-means clustering algorithm, which effectively segmented the EEG signals into distinct patterns based on their temporal and spatial characteristics across different frequency sub-bands. The methodology for extracting these features, along with the corresponding formulas, is explained in detail in Table 6. Each feature provides valuable insights into the dynamic properties of the EEG signals across multiple frequency domains and microstate configurations, enabling robust classification and comprehensive analysis with a substantial feature space that captures the complex spatiotemporal dynamics of neural oscillations.

Table 6: Feature Extraction Methods and Formulas for Microstate Analysis in EEG Data

| Feature | Description | Formula | Parameters Explanation |
|---|---|---|---|
| **Duration [28]** | The average time a microstate remains stable without switching. | $D = \frac{\sum_{i=1}^{N} t_i}{N}$ | $t_i$: Duration of microstate i, N: Total number of microstates in the signal. |
| **Occurrence [29]** | The number of times a specific microstate occurs per second. | $O = \frac{N}{T}$ | N: Total number of occurrences of the microstate, T: Total duration of the EEG signal. |
| **Coverage [29]** | Indicates the proportion of the total analysis time occupied by each microstate class. | $C = \frac{\sum_{i=1}^{N} t_i}{T} \times 100$ | $t_i$: Duration of microstate i, T: Total duration of the EEG signal. |
| **GFP (Global Field Power) [30]** | The spatial variance of the EEG signal at a specific time point. | $GFP(t) = \sqrt{\frac{1}{n} \sum_{i=1}^{n} (V_i(t) - \bar{V}(t))^2}$ | $V_i(t)$: Voltage at electrode iii at time t, $\bar{V}(t)$: Mean voltage at time t, n: Number of electrodes. |

Figure 3 shows an example of extracting 4-state, 5-state, 6-state, and 7-state microstates from a 64-channel EEG signal over a 30-second duration. A 10-second sliding window is applied to calculate the Global Field Power (GFP), and the k-means clustering method is used to identify the microstates (A-G). The topographical maps represent the distinct spatial patterns of brain activity for each microstate configuration. Unlike clustering-focused studies, we did not employ internal validation indices such as the silhouette coefficient, as our objective was to assess the discriminative value of microstate features through classification performance rather than optimizing cluster compactness.

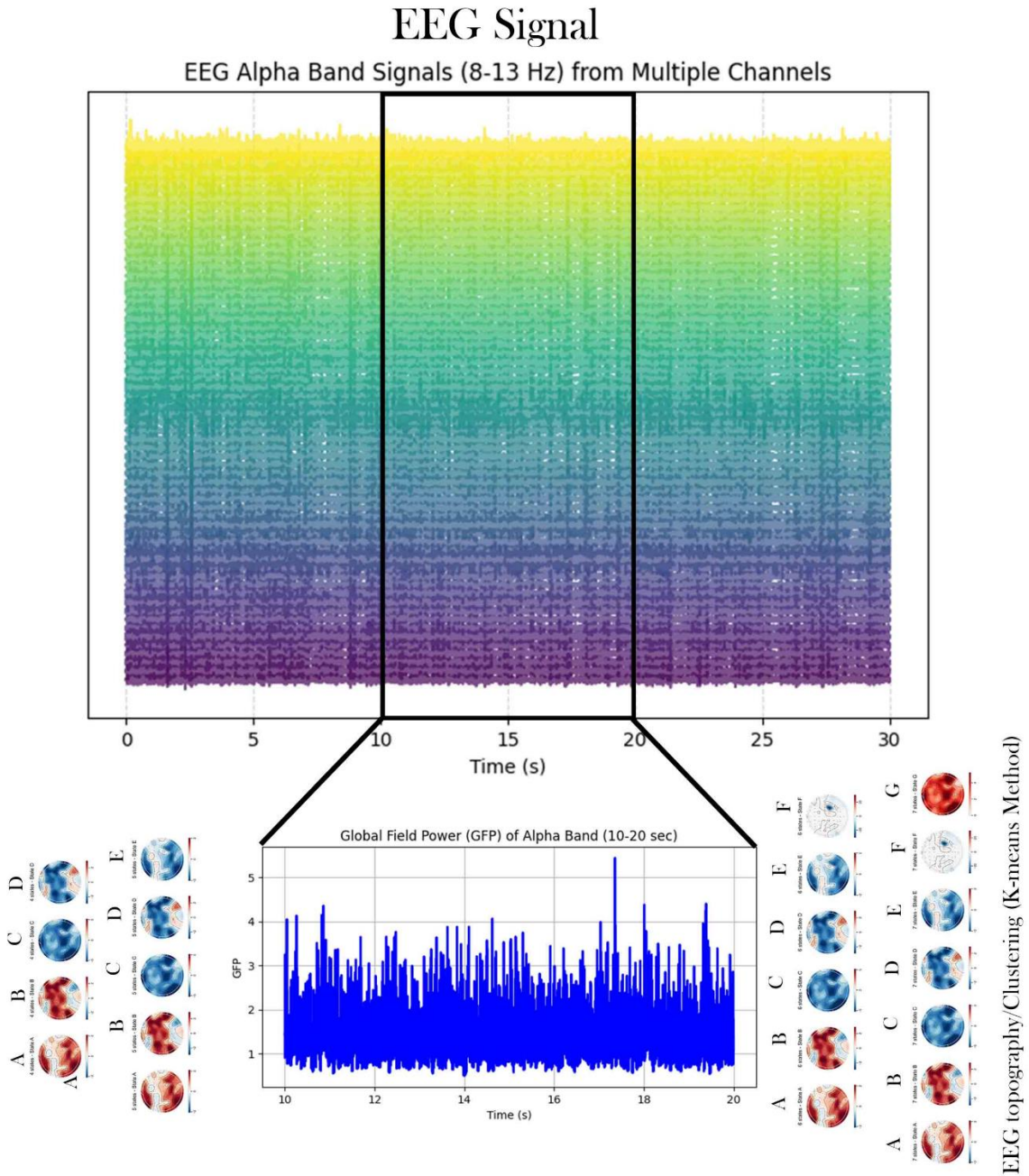

Figure 3. Sample of EEG Signal in the Alpha Band, Topographic Clustering, and Global Field Power (GFP)

### 2.5.1.2. Time-Frequency Imaging Using Continuous Wavelet Transform (CWT)

To enrich feature representation, each GFP signal extracted from the 64 EEG channels in all five frequency subbands was transformed into a 2D time-frequency image using Continuous Wavelet Transform (CWT). This process allowed the preservation of both temporal and frequency characteristics in a visual format suitable for CNN-based classification.

The transformation pipeline is summarized in Table 7, illustrating the key steps used in converting a GFP signal into a normalized 128×128 CWT image.

Table 7. Pipeline for generating CWT images from GFP signals extracted from EEG data.

| Step | Description |
| --- | --- |
| **GFP Extraction** | Computed from each of the 64 channels and 5 frequency bands |
| **Smoothing** | Moving average filter (window size = 5) |
| **Normalization** | Z-score normalization |
| **Wavelet Transform** | Continuous wavelet transform using cmor, with 127 scales |
| **Image Resizing** | Resized to 128×128 pixels and normalized for CNN input |

The method was applied consistently to all windows and participants. An example of these steps, applied to healthy and tinnitus subjects, is shown in Figure 4.

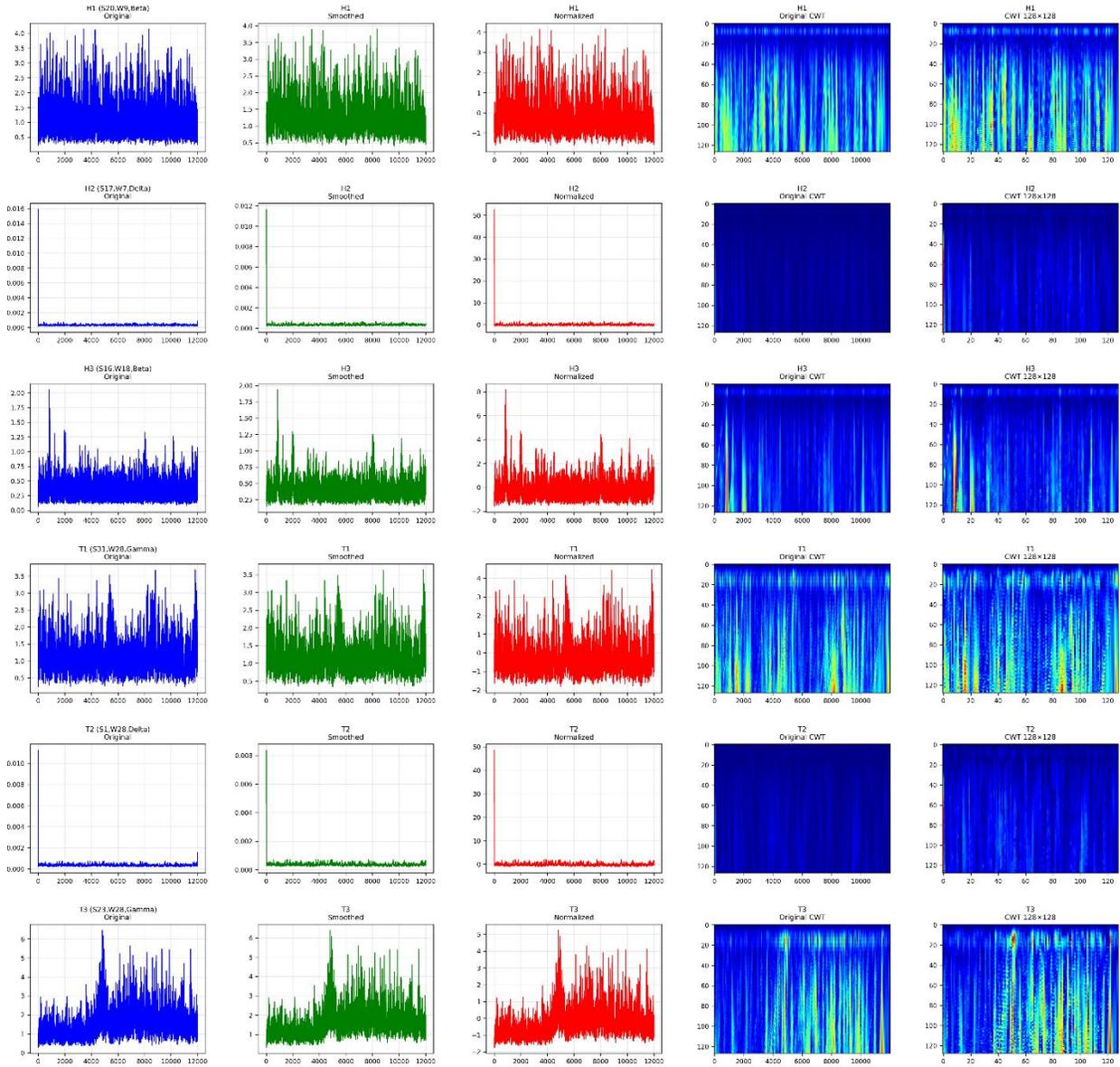

Figure 4. GFP to CWT transformation steps for representative EEG samples. Each row represents a different subject and frequency band. Columns show original GFP, smoothed signal, normalized signal, and the resulting 128×128 CWT image.

## 2.6. Classification Methodology

### 2.5.1. Deep Neural Network (DNN) Architecture

#### 2.5.1.1. DNN Model Design

DNNs are effective in modeling complex relationships in large datasets, particularly for classification tasks like medical image analysis and signal processing, by learning intricate patterns and nonlinearities. A typical DNN consists of an input layer, multiple hidden layers, and an output layer for classification. Regularization techniques, such as dropout and batch normalization, are used to enhance generalization and prevent overfitting [31]. The DNN model includes an input layer with N features, three hidden layers with 64, 32, and 16 neurons, and a single-output neuron for binary classification. The model employs the Adam optimizer, Binary Cross-Entropy loss, and regularization techniques including dropout (0.5), batch normalization, and L1/L2 kernel regularization. Table 8 summarizes the model architecture and parameters, and Figure 4 depicts the DNN architecture.

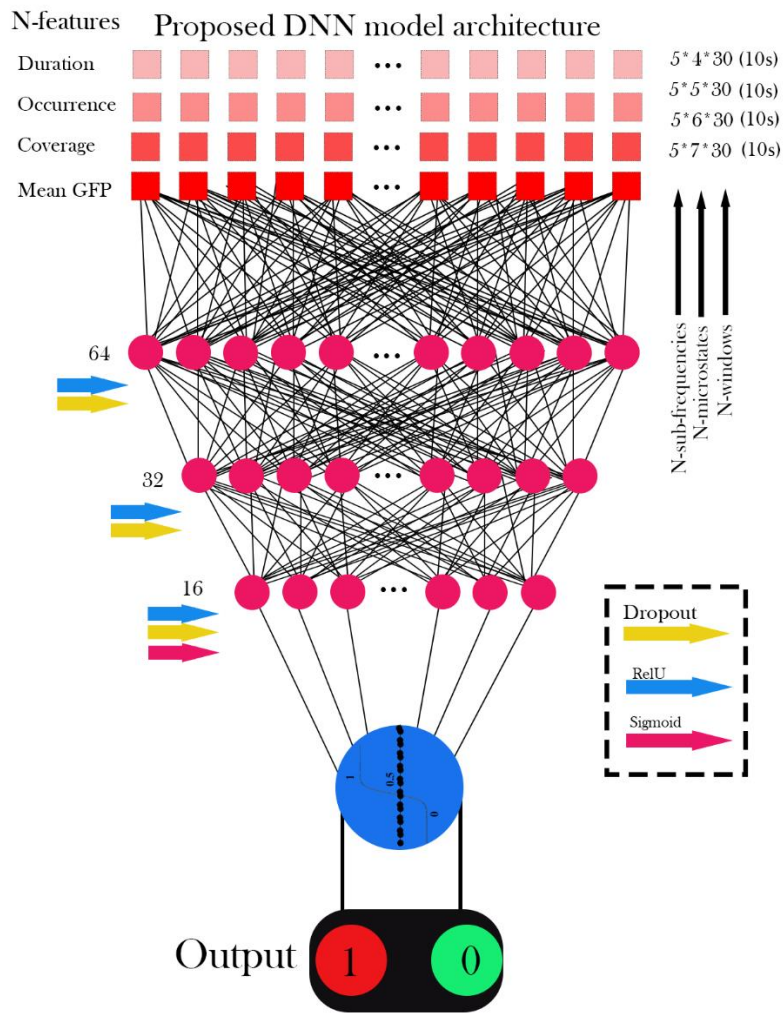

**Figure 4:** Workflow and Architecture of EEG Signal Classification Using DNN

## 2.6.2. Traditional Machine Learning Classifiers

### 2.6.2.1. Support Vector Machine (SVM)

SVM separates classes by maximizing the margin between the decision boundary and the nearest data points (support vectors) [32]:

$$f(x) = sign(\sum_{i=1}^{n} \alpha_i y_i K(x_i, x) + b) \tag{1}$$

Here, $x$ is the input feature vector, $x_i$ are the support vectors, $\alpha_i$ are the Lagrange multipliers, $y_i$ are class labels (+1 or 0), $K(x_i, x)$ is the kernel function (RBF kernel in this study), and $b$ is the bias term.

### 2.6.2.2. RF Model

RF combines multiple decision trees, with final prediction determined by majority voting [33]:

$$\hat{y} = mode\{T_1(x), T_2(x), \dots, T_m(x)\} \tag{2}$$

Here, $\hat{y}$ is the predicted class, $\dots, T_m(x)$ represents the prediction from the $m^{th}$ tree, and $x$ is the input feature vector.

### 2.6.2.3. DT Model

DT uses a sequence of decision rules based on impurity reduction using the Gini index [34]:

$$G = 1 - \sum_{t=1}^{k} p_i^2 \tag{3}$$

In this formula, $G$ is the Gini index, $p_i$ represents the proportion of samples in class $i$, and $k$ denotes the number of classes.

### 2.6.3. Convolutional Neural Network (CNN) Models

### 2.6.3.1. Independent CNN Model

The CNN model was designed to classify rs-fMRI images into healthy and tinnitus categories. The architecture consists of Conv2D layers with 3×3 filters and ReLU activation for feature extraction, MaxPooling layers for spatial down-sampling, and dense layers for classification. The output layer uses a sigmoid activation for binary classification. The model was optimized using the Adam optimizer with binary cross-entropy loss [31]. A batch size of 16, 10% validation split, and early stopping were employed to prevent overfitting. Dropout layers, batch normalization, and regularization techniques were used to enhance generalization. Table 8 summarizes the architecture and hyperparameters, while Figure 5 illustrates the CNN layers and connections used for classification tasks.

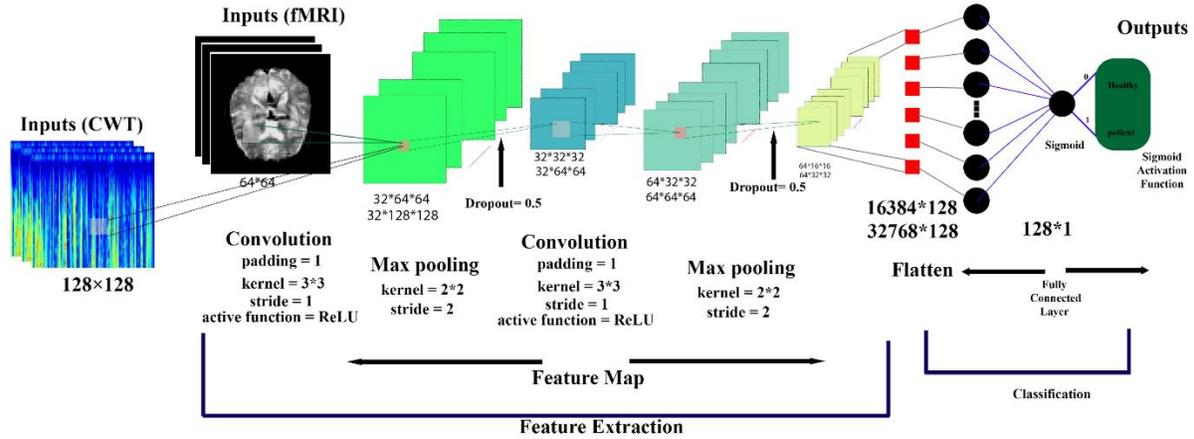

**Figure 5:** Architecture of a CNN Model

### 2.6.3.2. Pre-trained Architecture Implementation

The study employed VGG16 and ResNet50, two renowned pre-trained architectures, for the classification of rs-fMRI brain images into healthy and tinnitus categories. To ensure compatibility with the preprocessing steps applied to the rs-fMRI dataset, both models were fine-tuned. As the rs-fMRI images were initially grayscale, the single channel was replicated three times to meet the 3-channel (RGB) input specifications required by these pre-trained models.

**Optimizing VGG16 for Enhanced Performance:** VGG16 is a widely-used convolutional network with 16 layers pre-trained on ImageNet. It employs stacked convolutional layers with small filters and fully connected layers for classification. The last block is fine-tuned for this study **[35]**:

$$y = \sigma(W_{dense} \cdot GAP(h(X, W_{vgg}))) \tag{4}$$

Where $h(X, W_{vgg})$ refers to the feature extraction using the VGG16 layers, GAP represents Global Average Pooling, and σ is the Sigmoid function. X is the input image (64,64,3), $W_{vgg}$ are pre-trained VGG16 weights, and $W_{dense}$ denotes the final dense layer's weights. The learning parameters for fine-tuning VGG16 are detailed in Table 8. Figure 6 depicts the VGG16-based architecture for classifying rs-fMRI brain images.

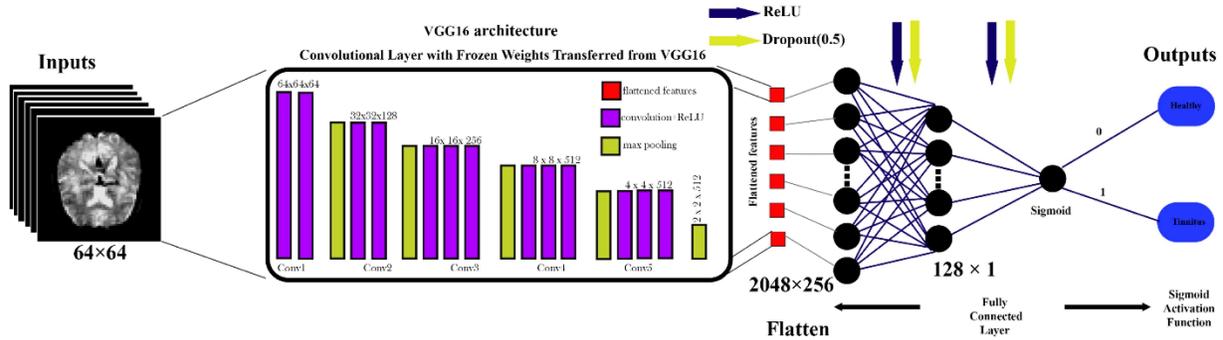

Figure 6: VGG16-Based Architecture for rs-fMRI Brain Image Classification.

**Optimizing ResNet50 for Enhanced Performance:** ResNet50 is a deep residual network pre-trained on ImageNet. It utilizes residual connections to ease the training of deep architectures by addressing the vanishing gradient problem. In this work, the last block of ResNet50 is fine-tuned for binary classification **[36]**:

$$y = \sigma(W_{dense} \cdot GAP(f(X, W_{res}))) \qquad (5)$$

Where $f(X, W_{res})$ is the residual feature map generated by ResNet50, GAP refers to Global Average Pooling, and σ is the Sigmoid activation. X is the input data (64,646,3), $W_{res}$ represents the pre-trained weights, and $W_{dense}$ corresponds to the final dense layer weights. Refer to Table 8 for the hyperparameters utilized for fine-tuning ResNet50. Figure 7 illustrates the ResNet-based architecture for classifying rs-fMRI brain images.

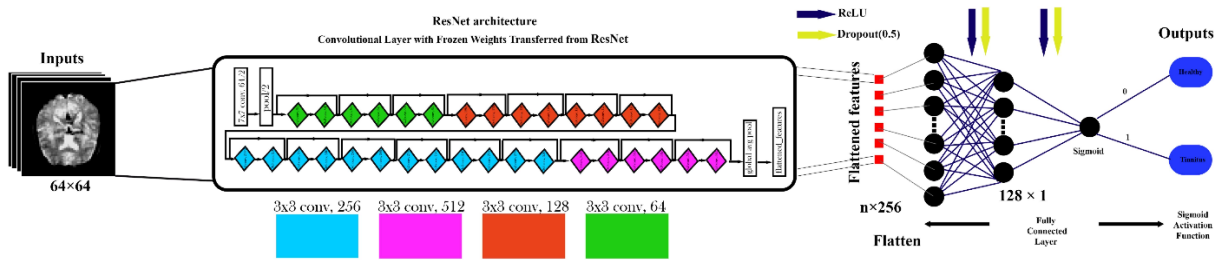

Figure 7: ResNet-Based Architecture for rs-fMRI Brain Image Classification.

Table 8 provides a comprehensive overview of the architecture designs, learning settings, and training configurations used for all implemented models in this study.

Table 8. Summary of Model Architectures, Learning Parameters, and Training Configurations Based on Classical and Deep Learning Approaches

| Model | Architecture Details | Learning/Training Parameters | Cross-Validation (K-fold = 5) | Activation Functions |
|---|---|---|---|---|
| **DNN** | Input: N features; Hidden layers: [64, 32, 16]; Output: 1 neuron | Optimizer: Adam; Loss: Binary Cross-Entropy; Batch size: 32; Epochs: 10; Dropout: 0.5; Batch Norm: Enabled; L1 = 0.005, L2 = 0.001 | Yes | ReLU in hidden layers, Sigmoid in output |
| **SVM** | Kernel-based classifier | Kernel: RBF; C=1.0; Gamma: scale; Standardization: Applied; Solver: SMO; Probability estimates: Enabled | Yes | Sigmoid |
| **DT** | Tree-based model with feature-based splitting | Splitting criterion: Gini impurity; Max depth: Unlimited; Min samples/split: 2; Standardization: Applied | Yes | Not applicable |
| **RF** | Ensemble of 100 decision trees with bootstrapping | Trees: 100; Criterion: Gini impurity; Max features: Auto; Standardization: Applied | Yes | Not applicable |
| **CNN** | Input: 64×64×1; Conv2D(32, 64 filters, 3×3, ReLU) + MaxPooling2D(2×2); Flatten → Dense(128, ReLU) → Dropout(0.5) → Output(1, Sigmoid) | Optimizer: Adam; Loss: Binary Cross-Entropy; Batch Size: 16; Epochs: 10; Dropout: 0.5; Batch Norm: True | Yes | ReLU in hidden layers, Sigmoid in output |
| **ResNet50, VGG16** | Input: 64×64×3; Base model: Pretrained on ImageNet (Frozen); Head: GlobalAvgPooling → Dense(256, ReLU) → Dropout(0.5) → Dense(128, ReLU) → Dropout(0.5) → Output(1, Sigmoid) | Optimizer: Adam; Loss: Binary Cross-Entropy; Learning rate: 0.0001; Batch size: 8; Epochs: 10; Early stopping (patience = 5) | Yes | ReLU in hidden layers, Sigmoid in output |

Figure 8 illustrates the proposed framework, encompassing preprocessing, signal decomposition, feature extraction, and classification of EEG signals alongside rs-fMRI data integration. The methodology utilizes CNN, DNN, and machine learning classifiers, incorporating pretrained models and a 5-fold cross-validation strategy for performance evaluation.

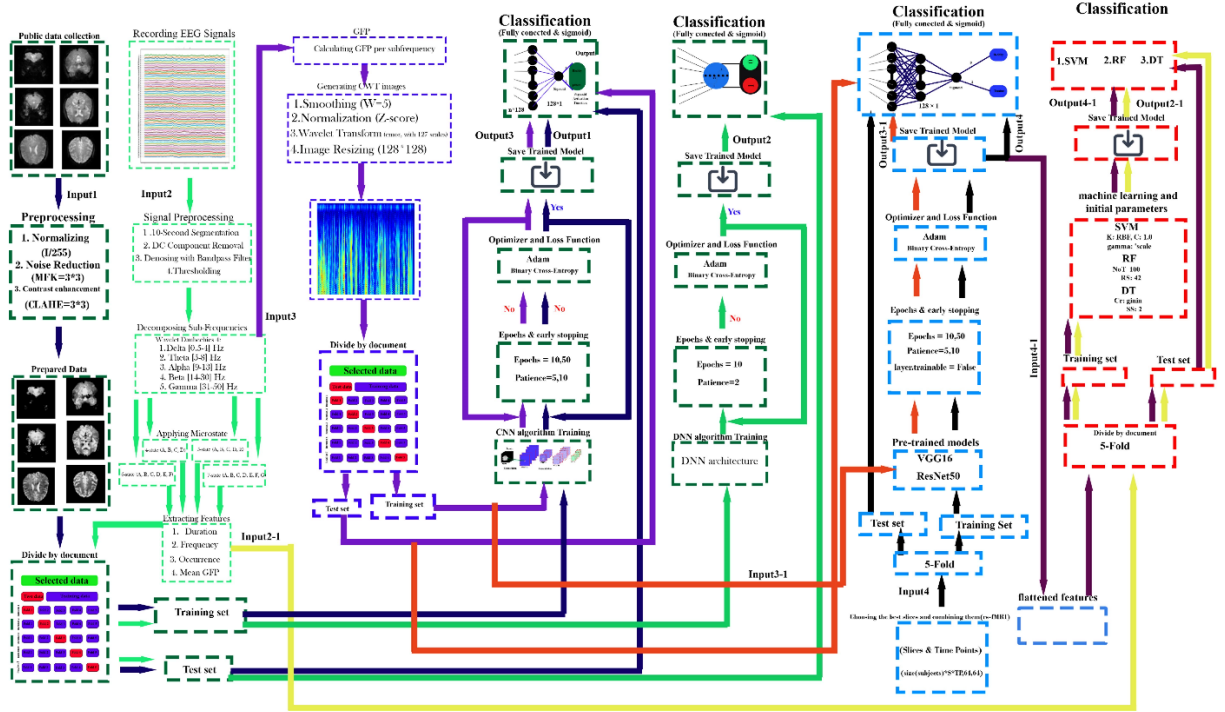

Figure 8: Proposed Framework for EEG Signal and rs-fMRI Image Processing with Multi-Model Classification

## 2.7. Model Evaluation and Validation

### 2.7.1 Performance Metrics

To assess model performance, we utilized Accuracy, Precision, Recall, F1 Score, ROC-AUC, and the Confusion Matrix. Accuracy measures the proportion of correctly classified samples, while Precision indicates the ratio of true positives to predicted positives. Recall, or sensitivity, evaluates the ability to identify all actual positives, and the F1 Score, as the harmonic mean of precision and recall, balances both metrics. ROC-AUC provides an aggregate measure of classification performance across thresholds, and the Confusion Matrix offers detailed insights into true/false positives and negatives [37].

$$Precision = \frac{TP}{TP + FP} \tag{6}$$

$$Accuracy = \frac{TP + TN}{TP + TN + FP + FN} \tag{7}$$

$$Recall = \frac{TP}{TP + FN} \tag{8}$$

$$F1\ Score = \frac{Precision \cdot Recall}{TPrecision + Recall} \tag{9}$$

$$TPR = \frac{TP}{TP+FN}, FPR = \frac{FP}{FP+TN} \tag{10}$$

### 2.7.2. Cross-Validation Strategy

We employed a rigorous subject-level 5-fold cross-validation strategy to ensure robust evaluation and prevent data leakage for both EEG and fMRI datasets. For the EEG dataset, the 80 participants (40 tinnitus patients, 40 healthy controls) were first stratified by clinical condition and systematically assigned to 5 folds, with each fold containing 16 participants (8 tinnitus, 8 healthy controls), ensuring that all 300 EEG windows from a single participant remain together within the same fold. Similarly, for the fMRI dataset, the 38 participants (19 tinnitus patients, 19 healthy controls) were stratified and assigned to 5 folds, with each fold containing 8 participants while maintaining class balance, ensuring that all brain slices and temporal sequences from individual subjects remain within the same fold. During each iteration, one fold serves as the independent test set while the remaining four folds constitute the training set, with this process repeated across all five folds for both modalities. This protocol strictly prevents data leakage by ensuring no participant appears in both training and testing sets simultaneously across either modality, guaranteeing that the models learn to classify clinical conditions rather than memorizing subject-specific patterns, thereby providing true generalization performance to unseen individuals and valid clinical condition discrimination.

## 3. Results

### 3.1. Dataset Characteristics and Preprocessing

In this study, 38 publicly available fMRI datasets with axial slices (64 × 64) were utilized, including 19 datasets from healthy individuals and 19 datasets from tinnitus patients recorded in a resting state. Additionally, EEG signals from 36 participants were recorded at rest using 64 channels, comprising 40 healthy individuals and 40 tinnitus patients. The preprocessing steps for the rs-fMRI data included median filtering, CLAHE, and normalization. For the EEG data, preprocessing involved 10-second segmentation, DC voltage removal, and the application of wavelet transform with the mother wavelet Daubechies 4 to extract sub-frequency bands: Delta (0.5–4 Hz), Theta (5–8 Hz), Alpha (9–13 Hz), Beta (14–30 Hz), and Gamma (31–50 Hz), followed by normalization. Moreover, microstate features extracted from the EEG signals, combining all channels, were calculated for 4, 5, 6, and 7-state microstates. These features included Duration, Occurrence, Mean Global Field Power (GFP), and Time Coverage.

## 3.2. CNN Performance Analysis on Resting-State fMRI Data

Figure 10 presents a comprehensive performance analysis of CNN models evaluated across 32 axial slices of rs-fMRI data, with results derived from subject-level 5-fold cross-validation. Each fold used 8 subjects (approximately 20% of 38 total subjects) for testing, resulting in 3,200 timepoint images per slice per fold for evaluation. Panel (a) displays the classification performance metrics including Precision, Accuracy, Recall, F1 Score, and ROC AUC across all slices, while panel (b) shows the corresponding average confusion matrices for each slice, comparing classification performance between Healthy and Tinnitus groups.

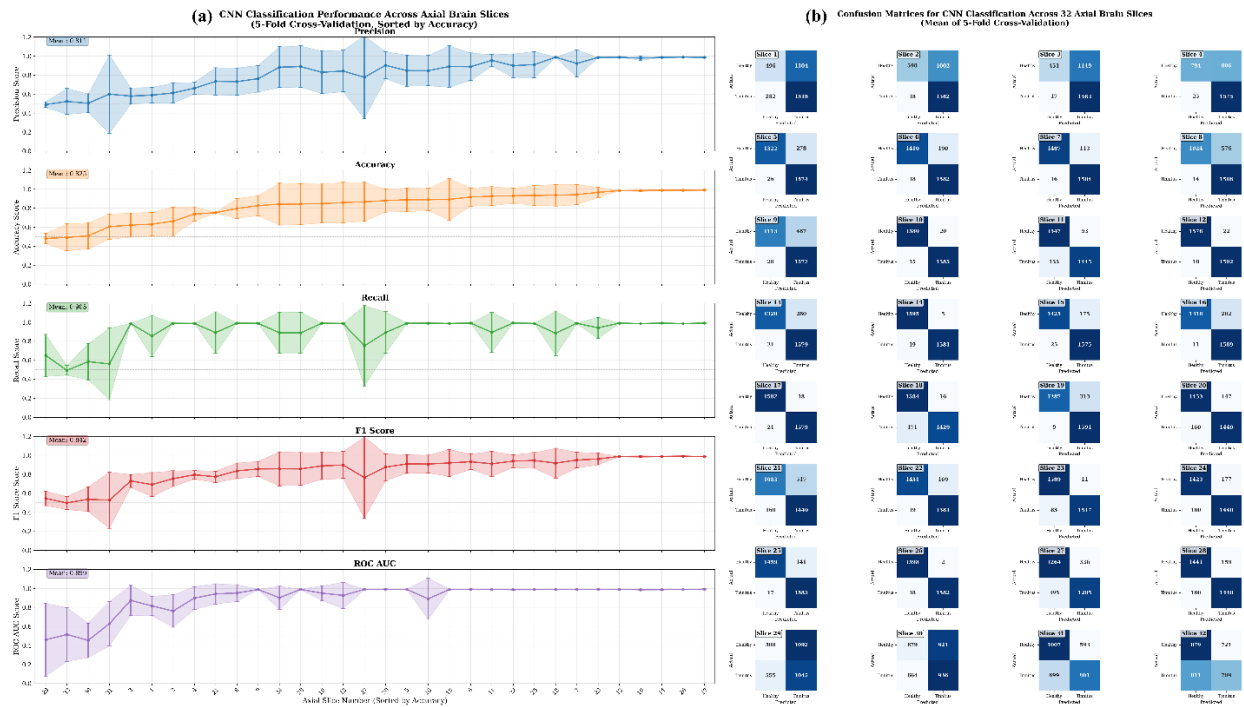

Figure 10: Comprehensive CNN Performance Analysis for rs-fMRI Data Classification Across 32 Axial Slices.

(a) CNN classification performance metrics (Precision, Accuracy, Recall, F1 Score, and ROC AUC) across 32 axial rs-fMRI slices, evaluated using subject-level 5-fold cross-validation. Each subject contributes 400 timepoint images per slice. The x-axis represents axial slice indices, sorted in ascending order by accuracy. Shaded areas indicate standard deviation across folds. (b) Mean confusion matrices (across 5 folds) for each axial slice, showing classification outcomes between healthy controls and tinnitus subjects at the timepoint level.

CNN-based analysis of 32 axial slices from resting-state fMRI data demonstrated significant spatial heterogeneity in discriminative capacity between tinnitus patients and healthy controls. The dataset comprised 19 subjects per group, with each subject contributing 400 timepoint images per axial slice for classification. Superior classification performance ($\geq$90% accuracy) was observed in slices 12, 10, 14, 26, 17, and ranks 28-32, with slice 17 achieving optimal discrimination (99.0% $\pm$ 0.4% accuracy, 99.2% $\pm$ 0.3% ROC AUC) and minimal misclassification (18 false positives, 21 false negatives; true positives: 1579, true negatives: 1382). High-performing slices 14, 12, and 26 exhibited comparably robust classification with false positive/negative counts below 30 timepoints each, while slice 14 demonstrated exceptional specificity with only 5 timepoints from healthy controls misclassified as tinnitus. Other high-performing slices include slice 26 with only 2 false positives and 18 false negatives (true positives: 1582, true negatives: 1398), and slice 10 with 20 false positives and 15 false negatives (true positives: 1585, true negatives: 1380). Conversely, inferior axial slices (29, 30, 31, 32) exhibited substantial classification errors, with slice 29 showing the poorest performance (508 false positives, 555 false negatives; true positives: 1045, true negatives: 1092) and slice 32 demonstrating comparable poor discrimination (721 false positives, 811 false negatives; true positives: 789, true negatives: 879). The dramatic performance gradient from these poorly performing slices to slice 17's near-perfect classification indicates that tinnitus-associated functional connectivity patterns are spatially localized to specific brain regions, suggesting that mid-to-superior axial slices contain the most diagnostically relevant neuroimaging biomarkers for automated tinnitus detection, likely corresponding to auditory processing regions and associated neural networks involved in tinnitus pathophysiology.

### 3.3. Neuroimaging-Based Comparative Analysis of Brain Activity Patterns

Figure 11 illustrates the comparison of axial fMRI slices between healthy individuals and tinnitus patients, highlighting the maximum pixel intensity differences across time points. For each subject in both groups, the first time point served as a reference, and the intensity of each voxel was compared with its corresponding voxel at subsequent time points. The maximum intensity difference for each voxel was calculated to capture the most prominent changes in neural activity over time. These maximum differences were then averaged across individuals within each group to generate representative slices for healthy and tinnitus subjects.

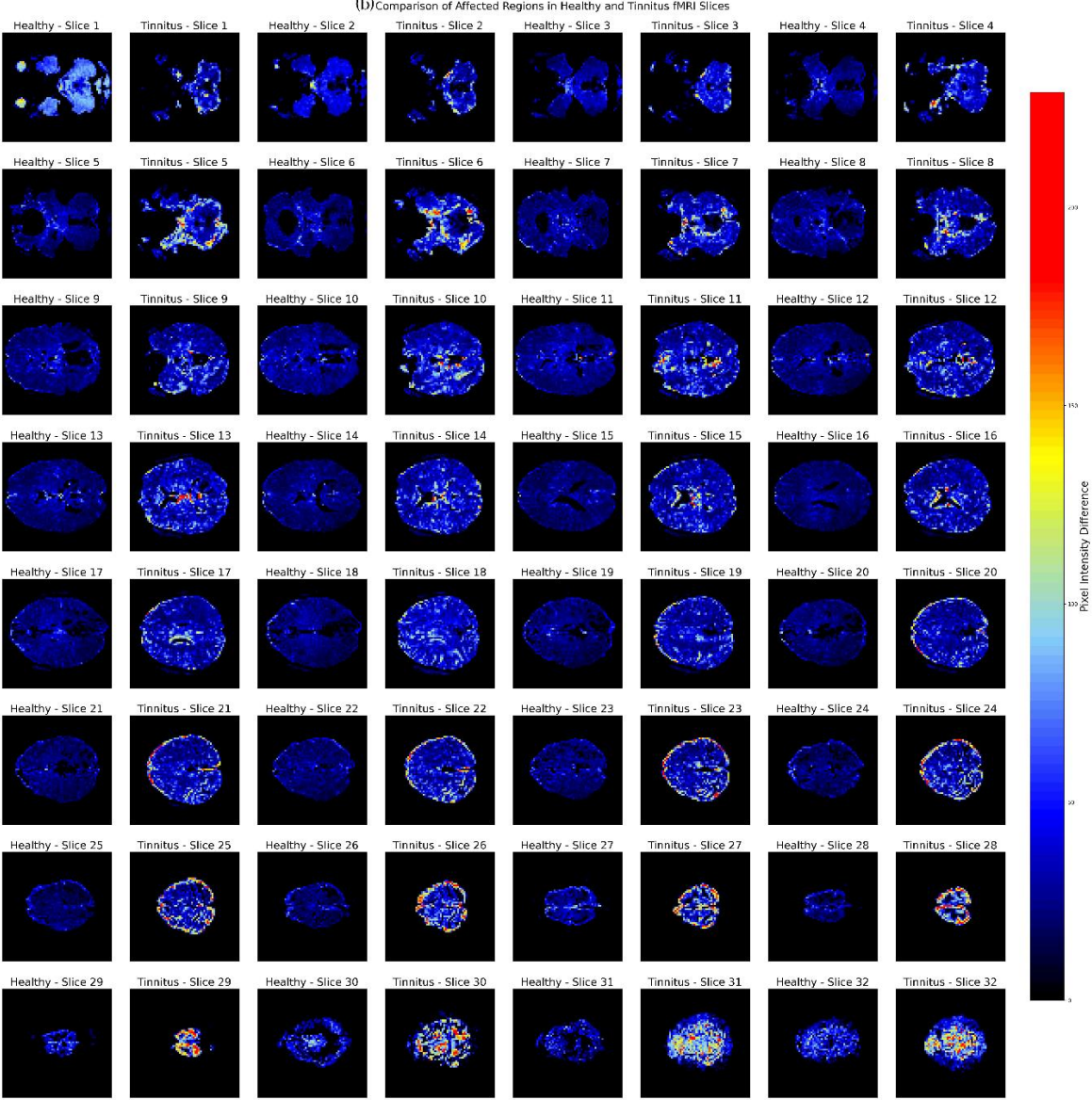

Figure 11: Axial fMRI Slice Comparison of Healthy and Tinnitus-Affected Brain Regions

In panel (a), the mean and standard deviation of each slice are displayed for both groups, with the x-axis representing the slice order and the y-axis showing the mean and standard deviation of pixel intensity differences in the final image. Panel (b) presents the maximum pixel intensity differences at different time points for healthy (odd-numbered columns from the left) and tinnitus (even-numbered columns from the left) groups. A customized color map highlights these differences, with red indicating higher intensity changes. The results reveal distinct patterns of neural activity and potentially affected regions in tinnitus patients. The color bar represents the pixel intensity difference scale.

The results indicate that tinnitus patients exhibit consistently higher mean pixel intensity differences across most fMRI slices compared to healthy individuals, particularly in slices 5–18 and 30–32. For instance, in slice 10, the mean intensity for tinnitus patients is 0.0776 compared to 0.0540 in healthy individuals, and in slice 13, it is 0.0812 versus 0.0477. The differences become more pronounced in deeper slices, with slice 31 showing a mean of 0.0721 for tinnitus patients, significantly higher than 0.0162 in healthy individuals. Additionally, the standard deviation is notably higher in the tinnitus group, reaching 0.1572 in slice 31 compared to 0.0522 in the healthy group, suggesting greater variability in neural activity. These findings highlight distinct alterations in brain activity patterns in tinnitus patients, potentially reflecting abnormal neural dynamics associated with the condition.

### 3.4. fMRI-Based Classification Performance: Pre-trained and Hybrid Model Evaluation

Figure 12 demonstrates the classification performance of combined high-performing fMRI slices from two groups (healthy controls and tinnitus patients) using eight different classifiers: VGG16, VGG16-DT (Decision Tree), VGG16-RF (Random Forest), VGG16-SVM (Support Vector Machine), ResNet50, ResNet50-DT, ResNet50-RF, and ResNet50-SVM across five evaluation metrics. Based on the individual slice analysis, slices achieving accuracy above 90% (slices 6, 7, 10, 11, 12, 14, 17, 18, 22, 23, 25, 26) were selected and combined to optimize classification performance. Results represent the mean performance of 5-fold cross-validation on test data.

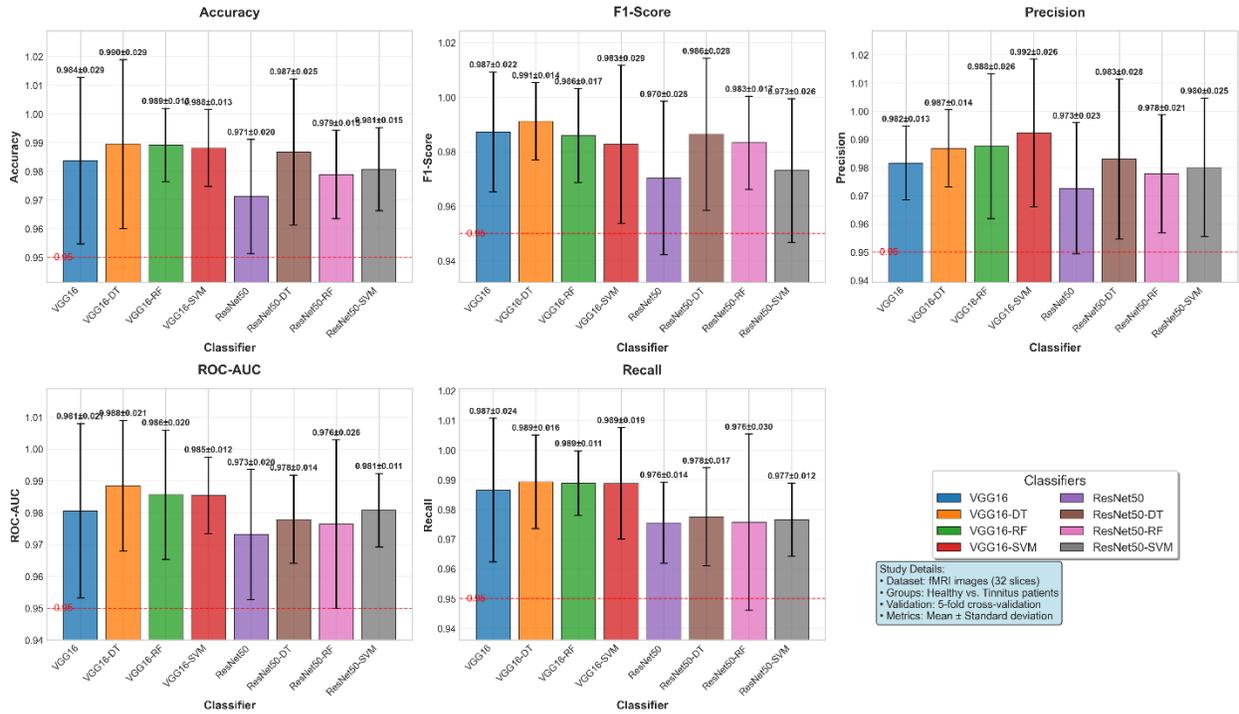

Figure 12. Performance Comparison of Pre-trained and Hybrid Models for fMRI-based Tinnitus Classification. Data represent mean ± standard deviation from 5-fold cross-validation on test data. The red dashed line indicates the 95% clinical relevance threshold. Hybrid models combine pre-trained CNN feature extraction with traditional machine learning classifiers (Decision Tree, Random Forest, Support Vector Machine) for automated tinnitus detection.

The classification of 32-slice fMRI images from healthy controls and tinnitus patients demonstrated exceptional accuracy performance across all evaluated models, with scores ranging from 97.12% (ResNet50) to 98.95% (VGG16-DT). Hybrid approaches consistently achieved superior accuracy compared to standalone CNN architectures, with VGG16-based models substantially outperforming ResNet50 variants (average accuracy: 98.76% vs. 97.94%, respectively). VGG16-DT emerged as the optimal classifier, demonstrating the highest accuracy (98.95% ± 2.94%) and ROC-AUC (98.84% ± 2.05%), followed closely by VGG16-RF with 98.91% ± 1.28% accuracy and VGG16-SVM with 98.81% ± 1.34% accuracy. ROC-AUC values consistently exceeded 97% across all models, ranging from 97.31% (ResNet50) to 98.84% (VGG16-DT), indicating excellent discriminative ability between healthy and tinnitus groups. The most stable accuracy performance was achieved by VGG16-RF (±1.28% standard deviation), while VGG16-SVM showed the most consistent ROC-AUC (98.54% ± 1.20%). All models significantly exceeded the 95% clinical relevance threshold for both accuracy and ROC-AUC metrics, demonstrating robust diagnostic capability suitable for automated clinical tinnitus detection. The integration of traditional machine learning classifiers with pre-trained CNN feature extraction proved highly effective, with hybrid approaches improving accuracy by an average of 1.2% for VGG16 variants and 2.8% for ResNet50 variants compared to their standalone counterparts.

### 3.5. EEG Microstate Analysis and Statistical Characterization

Figure 13 provides a detailed statistical analysis of EEG microstate features across different microstate configurations and frequency bands in both healthy and tinnitus groups. The microstate configurations include the 4-state (A, B, C, D), 5-state (A, B, C, D, E), 6-state (A, B, C, D, E, F), and 7-state (A, B, C, D, E, F, G) cluster solutions. Frequency bands considered are delta (1–4 Hz), theta (4–8 Hz), alpha (8–12 Hz), beta (13–30 Hz), and gamma (30–45 Hz). The extracted features include duration (ms), coverage (%), occurrence (events per epoch), and mean global field power (GFP). Table 9 summarizes the top 15 microstate-frequency-feature combinations based on the highest effect sizes (Cohen's d) and statistical power.

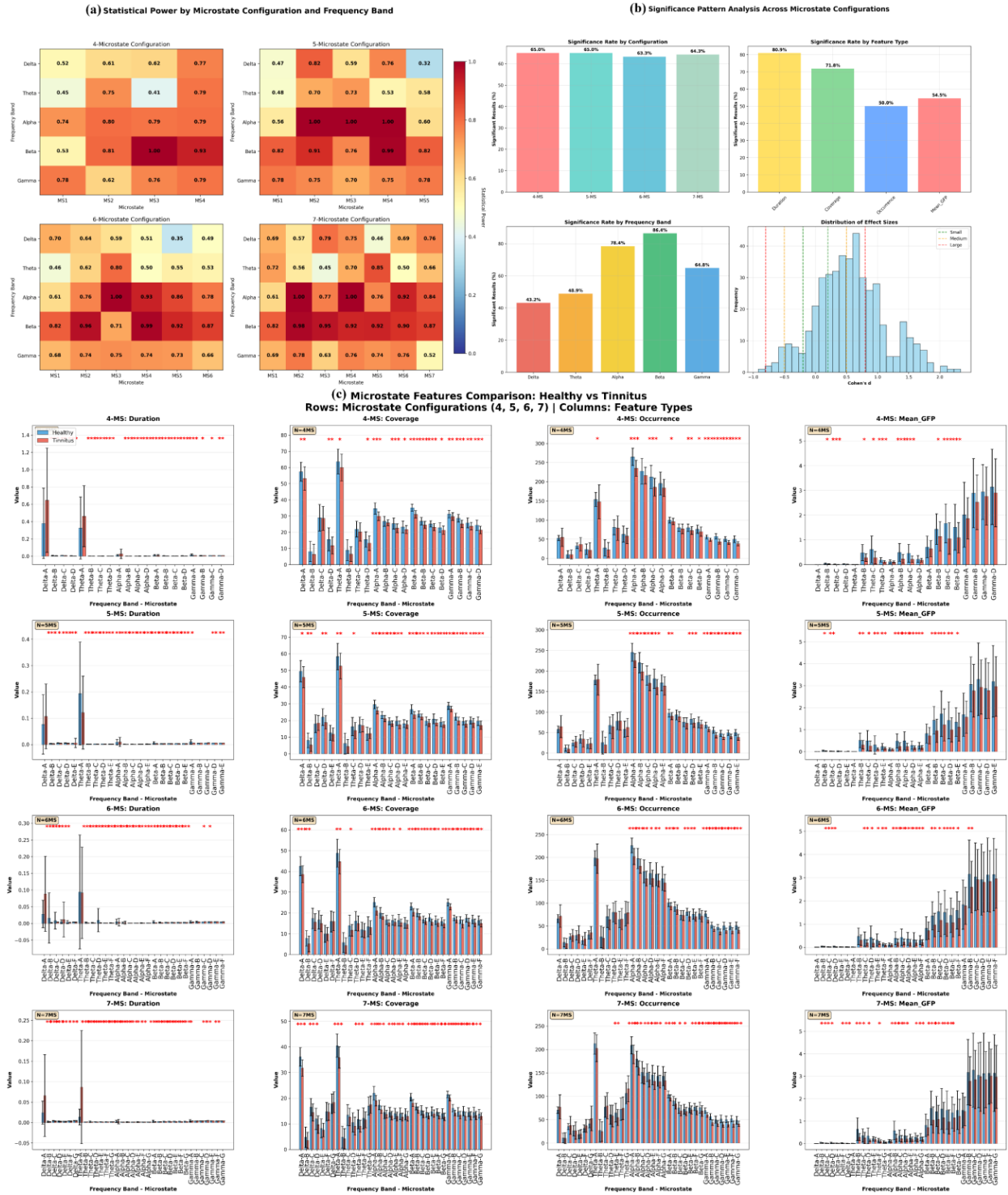

Figure 13. Comprehensive Statistical and Visual Comparison of Microstate Features Across Configurations and Frequency Bands in Healthy vs. Tinnitus Groups: Statistical Power (a) Significance Patterns (b) Microstate Feature Comparison (c).

Subfigure (a) illustrates the statistical power across microstate configurations and frequency bands, highlighting the varying effect strengths. Subfigure (b) summarizes significance rates across configurations, feature types, and frequency bands, along with the distribution of effect sizes. Subfigure (c) compares five key microstate features (duration, coverage, occurrence, and mean GFP) between groups across all configurations and frequency bands. Statistically significant group differences (p < 0.05) are marked with red asterisks.

Table 9: Top 15 EEG Microstate Findings: Healthy vs Tinnitus

| Rank | Cluster Solution | Frequency Band | Microstate | Feature | Healthy Mean | Tinnitus Mean | Cohen's d | P-value | Statistical Power |
|---|---|---|---|---|---|---|---|---|---|
| 1 | 7-state | Gamma | B | Occurrence (events/epoch) | 56.56 | 43.81 | 2.11 | $1.58 \times 10^{-14}$ | 1.000 |
| 2 | 4-state | Alpha | A | Coverage (%) | 34.62% | 29.95% | 1.48 | $3.80 \times 10^{-9}$ | 1.000 |
| 3 | 5-state | Alpha | B | Duration (ms) | 1.23 ms | 1.11 ms | 1.89 | $1.25 \times 10^{-12}$ | 1.000 |
| 4 | 6-state | Beta | A | Coverage (%) | 23.09% | 20.28% | 1.58 | $6.06 \times 10^{-10}$ | 1.000 |
| 5 | 4-state | Beta | A | Coverage (%) | 35.14% | 31.06% | 1.81 | $6.59 \times 10^{-12}$ | 1.000 |
| 6 | 7-state | Alpha | B | Duration (ms) | 1.12 ms | 0.99 ms | 2.20 | $5.40 \times 10^{-14}$ | 1.000 |
| 7 | 6-state | Gamma | A | Occurrence (events/epoch) | 76.79 | 62.95 | 2.34 | $1.77 \times 10^{-16}$ | 1.000 |
| 8 | 4-state | Alpha | A | Occurrence (events/epoch) | 264.81 | 234.90 | 1.39 | $2.32 \times 10^{-8}$ | 1.000 |
| 9 | 5-state | Gamma | A | Occurrence (events/epoch) | 68.17 | 57.52 | 1.70 | $5.79 \times 10^{-11}$ | 1.000 |
| 10 | 7-state | Gamma | A | Occurrence (events/epoch) | 69.65 | 60.32 | 1.75 | $1.89 \times 10^{-11}$ | 1.000 |
| 11 | 4-state | Theta | D | Mean GFP (μV) | 0.191 | 0.113 | 1.08 | $9.67 \times 10^{-6}$ | 1.000 |
| 12 | 6-state | Alpha | A | Coverage (%) | 25.16% | 21.09% | 1.86 | $2.11 \times 10^{-12}$ | 1.000 |
| 13 | 5-state | Beta | D | Duration (ms) | 3.31 ms | 2.93 ms | 1.54 | $1.28 \times 10^{-9}$ | 1.000 |
| 14 | 7-state | Beta | D | Duration (ms) | 2.70 ms | 2.34 ms | 1.72 | $3.78 \times 10^{-11}$ | 1.000 |
| 15 | 6-state | Gamma | B | Occurrence (events/epoch) | 51.82 | 43.51 | 1.40 | $1.89 \times 10^{-8}$ | 1.000 |

The comparative analysis between healthy controls and tinnitus patients revealed systematic disruptions in EEG microstate dynamics across multiple frequency bands and clustering configurations. The most pronounced alterations were observed in gamma-band microstates, with microstate B occurrence rates in the 7-state configuration demonstrating the largest effect size (Cohen's d = 2.11, p = $1.58 \times 10^{-14}$), where healthy participants exhibited significantly higher rates (56.56 vs 43.81 events/epoch). Gamma-band microstate A consistently showed reduced occurrence rates across all clustering solutions (5-state: d = 1.70; 6-state: d = 2.34; 7-state: d = 1.75). Alpha-band microstates revealed systematic reductions in tinnitus patients, with microstate A coverage decreased in both 4-state (34.62% vs 29.95%, d = 1.48) and 6-state (25.16% vs 21.09%, d = 1.86) configurations, alongside reduced occurrence rates (4-state: 264.81 vs 234.90 events/epoch, d = 1.39) and shortened microstate B durations in 5-state (1.23 vs 1.11 ms, d = 1.89) and 7-state (1.12 vs 0.99 ms, d = 2.20) analyses. Beta-band analysis demonstrated consistent alterations including reduced microstate A coverage across 4-state (35.14% vs 31.06%, d = 1.81) and 6-state (23.09% vs 20.28%, d = 1.58) configurations, and shortened microstate D durations in both 5-state (3.31 vs 2.93 ms, d = 1.54) and 7-state (2.70 vs 2.34 ms, d = 1.72) clustering solutions. Additionally, theta-band microstate D exhibited significantly reduced mean global field power in tinnitus patients (0.191 vs 0.113 μV, d = 1.08, p = $9.67 \times 10^{-6}$). These findings collectively demonstrate that tinnitus is characterized by widespread disruptions in neural network dynamics, particularly affecting high-frequency gamma oscillations and alpha-band resting-state networks, with all significant results exhibiting exceptionally high statistical power (>0.99) and effect sizes ranging from medium to very large (Cohen's d = 1.08-2.34), indicating robust and clinically meaningful differences that potentially reflect the underlying pathophysiological mechanisms of phantom auditory perception.

### 3.6. Machine Learning Classification Performance on EEG Microstate Features

#### 3.6.1. Comparative Performance Analysis

Figure 14 illustrates the performance of the Deep Neural Network (DNN), Random Forest (RF), Decision Tree (DT), and Support Vector Machine (SVM) models for classifying EEG signals using all extracted features, including Duration, Occurrence, Mean Global Field Power (GFP), and Time Coverage, from 4-state, 5-state, 6-state, and 7-state microstates across the five sub-frequency bands: Delta, Theta, Alpha, Beta, and Gamma. All reported results represent the mean across 5 folds for the test data.

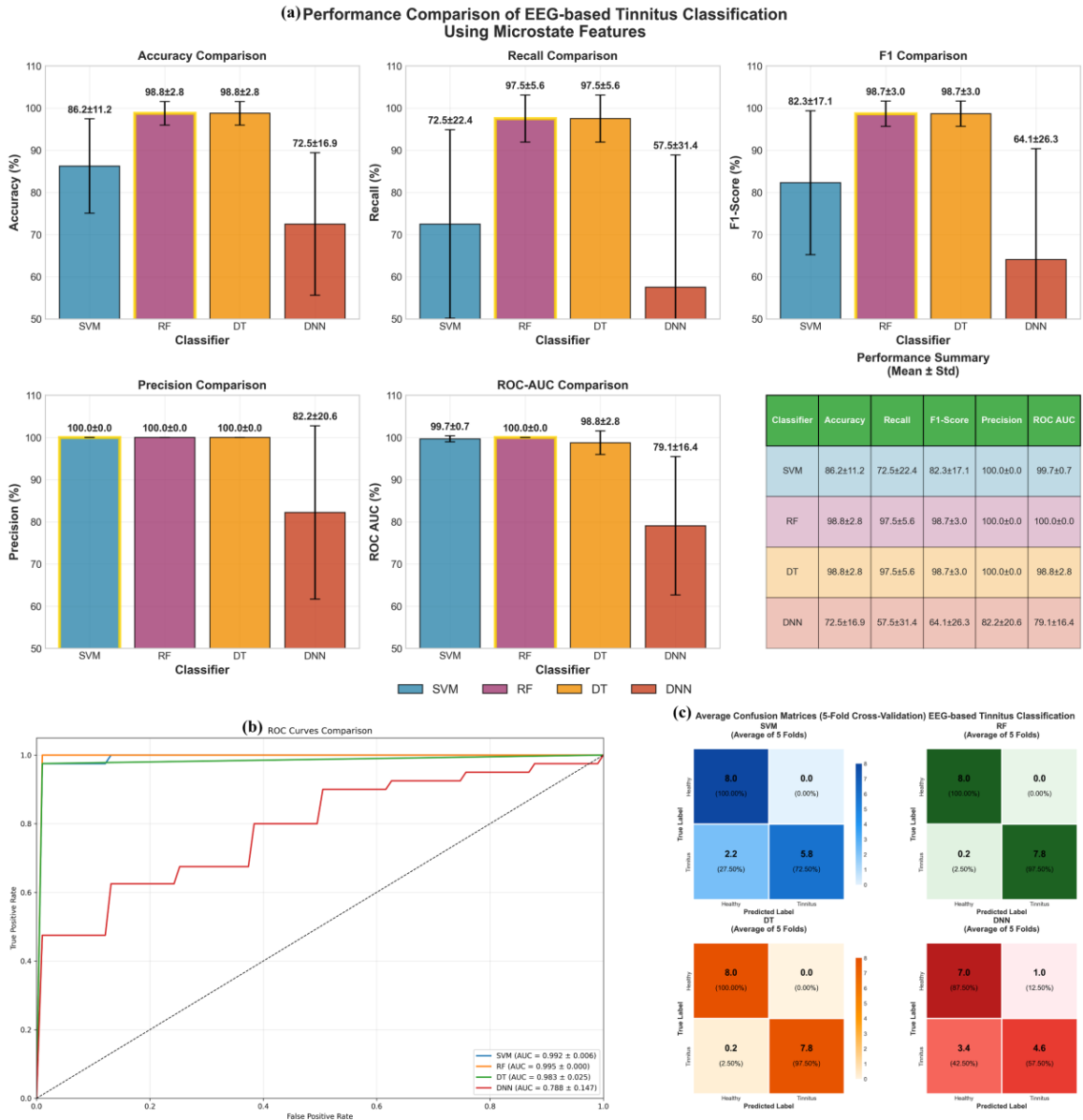

Figure 14. Performance Evaluation of Machine Learning Classifiers for EEG-based Tinnitus Classification Using Comprehensive Microstate Features

Panel (a) presents bar charts comparing five key performance metrics across the four algorithms with error bars indicating variability, alongside a numerical summary table. Panel (b) displays ROC curves that demonstrate the discrimination capability of each classifier against the diagonal reference line. Panel (c) shows 2×2 confusion matrices for each classifier, presenting average classification results with both absolute counts and percentage distributions.

The evaluation reveals distinct performance characteristics among the tested classifiers. Both Random Forest (RF) and Decision Tree (DT) algorithms demonstrated consistently superior performance, each achieving an accuracy of 98.8 ± 2.8% and perfect precision (100.0%), with a recall of 97.5 ± 5.6%, indicating reliable sensitivity in detecting tinnitus cases. The Support Vector Machine (SVM) showed moderate classification capability, yielding an accuracy of 86.2 ± 11.2% with similarly perfect precision (100.0 ± 0.0%) but notably lower recall (72.5 ± 22.4%), suggesting a tendency to miss some true positives. In contrast, the Deep Neural Network (DNN) exhibited the most variable and suboptimal performance across all metrics, with an accuracy of 72.5 ± 16.9% and considerable standard deviations, reflecting instability across the cross-validation folds. The ROC analysis confirmed these trends, with RF and DT achieving ROC-AUC scores of 98.8 ± 2.8%, while DNN produced the lowest area under the curve at 79.1 ± 16.4%. Examination of the confusion matrices revealed that classification errors were predominantly false negatives rather than false positives, particularly for the DNN model, where tinnitus cases were frequently misclassified as healthy controls. These findings underscore the robustness and reliability of tree-based models, particularly RF and DT, for classifying EEG-derived microstate features associated with tinnitus.

### 3.6.2. Statistical Validation Using DeLong Tes

To statistically validate the differences in ROC-AUC values between classifiers, pairwise comparisons were conducted using the DeLong test, as shown in Table 10.

Table 10. Pairwise ROC-AUC Comparisons Using the DeLong Test Between Classifiers

| Classifier 1 | Classifier 2 | $AUC_1$ | $AUC_2$ | Z-statistic | P-value | Significant |
|---|---|---|---|---|---|---|
| **SVM** | RF | 0.9969 | 1.0000 | -0.0828 | 0.9340 | No |
| **SVM** | DT | 0.9969 | 0.9875 | 0.2518 | 0.8012 | No |
| **SVM** | DNN | 0.9969 | 0.8125 | 3.1363 | 0.0017 | Yes |
| **RF** | DT | 1.0000 | 0.9875 | 0.3601 | 0.7188 | No |
| **RF** | DNN | 1.0000 | 0.8125 | 3.1354 | 0.0017 | Yes |
| DT | DNN | 0.9875 | 0.8125 | 2.9327 | 0.0034 | Yes |

The results revealed that the differences between RF, DT, and SVM were not statistically significant (p > 0.05), indicating comparable discriminative capabilities among these models. However, the DNN model was significantly inferior in performance compared to all other classifiers (p < 0.01), confirming its limited ability to distinguish between healthy and tinnitus subjects. These findings underscore the robustness of tree-based models (RF and DT) for EEG-based microstate classification and highlight their advantage over deep learning methods in scenarios with limited training data and high feature complexity.

### 3.6.3. Comprehensive Performance Evaluation Across Multiple Experimental Conditions

Figure 15 shows the classification results for distinguishing between Healthy and Tinnitus groups based on four microstate features (Duration, Occurrence, Mean GFP, and Coverage). The evaluations were conducted across five EEG sub-bands (delta, theta, alpha, beta, gamma) and four microstate configurations (4-state, 5-state, 6-state, and 7-state). For each configuration, combinations of one to four features were analyzed using four classifiers: SVM, RF, DT, and DNN. The reported values represent the average performance across five-fold cross-validation on the test data.

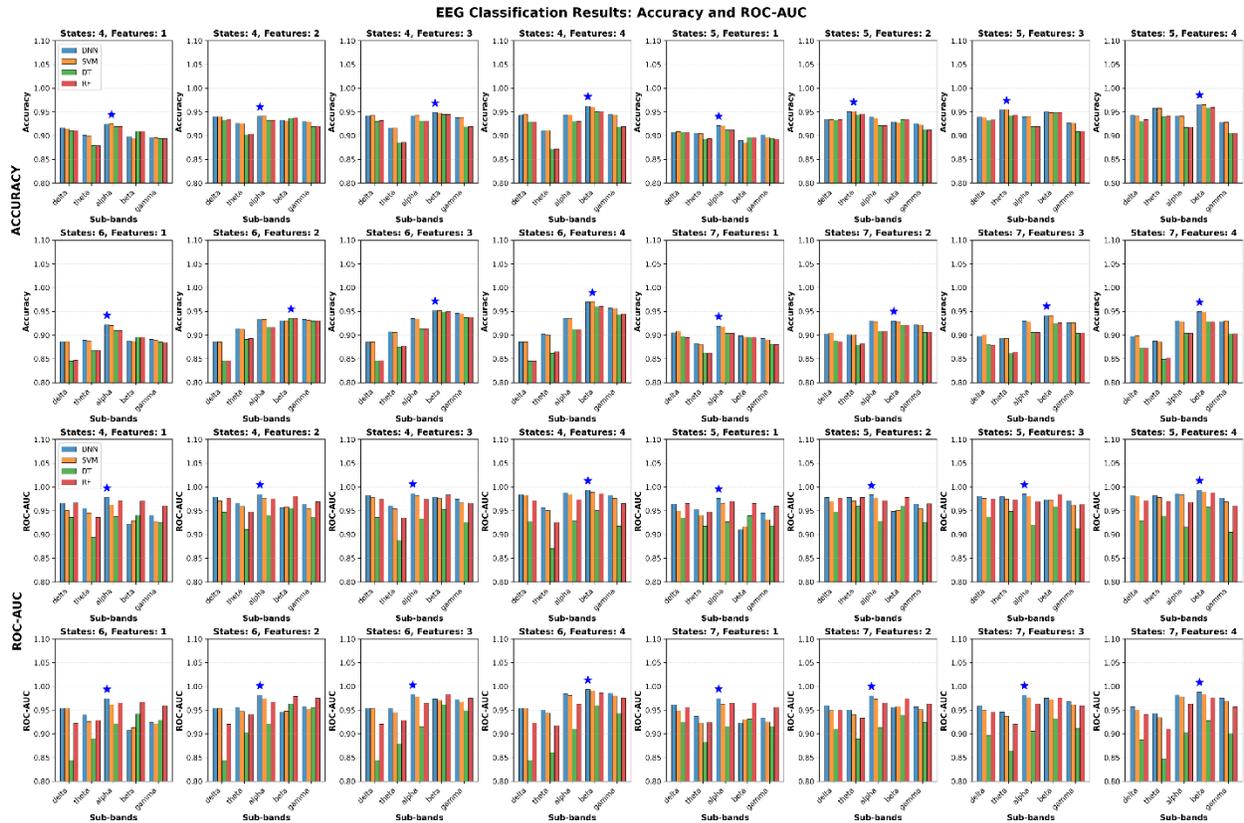

Figure 15. EEG-Based Classification Performance Across Microstate Configurations and Frequency Bands.

Each subplot illustrates the classification performance (top: Accuracy; bottom: ROC-AUC) for a specific combination of microstate configuration and number of features used. The x-axis denotes the EEG sub-bands, and the bar groups represent results from different classifiers. The blue star in each subplot indicates the highest achieved metric value across all conditions.

Tables 11-14 provide a comprehensive analysis of the EEG-based classification results for distinguishing between healthy controls and tinnitus patients. Table 11 presents the optimal performance achieved by each classifier across all experimental conditions, while Table 12 compares the discriminative power of different EEG frequency sub-bands. Table 13 evaluates the impact of microstate configuration complexity on classification accuracy, and Table 14 analyzes how feature dimensionality affects overall performance metrics.

Table 11: Best Performance per Classifier (Accuracy & ROC-AUC)

| Classifier | Sub-band | States | Features | Best Accuracy | Best ROC-AUC |
|---|---|---|---|---|---|

| Classifier | Sub-band | | | Accuracy | ROC-AUC |
|---|---|---|---|---|---|
| DNN | Beta | 6 | 4 | 0.9696 | 0.9932 |
| SVM | Beta | 6 | 4 | 0.9698 | 0.9905 |
| DT | Beta | 6 | 4 | 0.9600 | 0.9622 |
| RF | Beta | 6 | 4 | 0.9602 | 0.9874 |

Table 12: Sub-band Performance Analysis (Mean Performance Across All Configurations)

| Sub-band | Mean Accuracy ± SD | Mean ROC-AUC ± SD |
|---|---|---|
| Delta | 0.9063 ± 0.0234 | 0.9447 ± 0.0287 |
| Theta | 0.9023 ± 0.0307 | 0.9353 ± 0.0349 |
| Alpha | 0.9256 ± 0.0119 | 0.9565 ± 0.0221 |
| Beta | 0.9308 ± 0.0246 | 0.9607 ± 0.0267 |
| Gamma | 0.9158 ± 0.0205 | 0.9458 ± 0.0237 |

Table 13: Micro-state Configuration Performance (Mean Performance Across All Sub-bands and Classifiers)

| States | Mean Accuracy ± SD | Mean ROC-AUC ± SD |
|---|---|---|
| 4 | 0.9196 ± 0.0209 | 0.9549 ± 0.0254 |
| 5 | 0.9214 ± 0.0215 | 0.9567 ± 0.0251 |
| 6 | 0.9148 ± 0.0323 | 0.9478 ± 0.0349 |
| 7 | 0.9004 ± 0.0198 | 0.9386 ± 0.0252 |

Table 14: Feature Combination Performance (Mean Performance Across All Sub-bands and Classifiers)

| Features | Mean Accuracy ± SD | Mean ROC-AUC ± SD |
|---|---|---|
| 1 | 0.8942 ± 0.0129 | 0.9377 ± 0.0194 |
| 2 | 0.9181 ± 0.0205 | 0.9518 ± 0.0213 |
| 3 | 0.9223 ± 0.0235 | 0.9547 ± 0.0262 |
| 4 | 0.9245 ± 0.0269 | 0.9574 ± 0.0297 |

The experimental results reveal several important patterns in EEG classification performance. Table 11 demonstrates that SVM achieved the highest accuracy (96.98%) while DNN obtained the best ROC-AUC score (99.32%), both in the beta sub-band with optimal configurations. The sub-band analysis in Table 12 indicates that beta frequency consistently outperformed other bands with mean accuracy of 93.08% ± 2.46% and mean ROC-AUC of 96.07% ± 2.67%, followed by alpha and delta sub-bands. Table 13 shows that 5 micro-states provided the most balanced performance with mean ROC-AUC of 95.67% ± 2.51%, while 6 micro-states yielded the highest individual accuracy but with greater variability (91.48% ± 3.23%). The feature combination analysis in Table 14 demonstrates a progressive improvement with increased feature count, where 4-feature combinations achieved the best mean performance (accuracy: 92.45% ± 2.69%, ROC-AUC: 95.74% ± 2.97%) compared to single-feature approaches.

### 3.7. Deep Learning Analysis of CWT-Transformed EEG Signals

#### 3.7.1. Frequency-Specific Performance Patterns

Figure 16 illustrates the comprehensive performance evaluation of three deep learning architectures (CNN, fine-tuned ResNet50, and fine-tuned VGG16) applied to CWT-transformed GFP features extracted from five EEG frequency sub-bands (Delta, Theta, Alpha, Beta, Gamma). The figure displays radar charts comparing five classification metrics across frequency bands, ROC curves demonstrating the discriminative capability of each model, and detailed confusion matrices showing the classification outcomes for healthy controls versus tinnitus patients. The visualization provides a multi-dimensional analysis of model performance using 5-fold cross-validation results on 64-channel EEG data transformed into 128×128 CWT images.

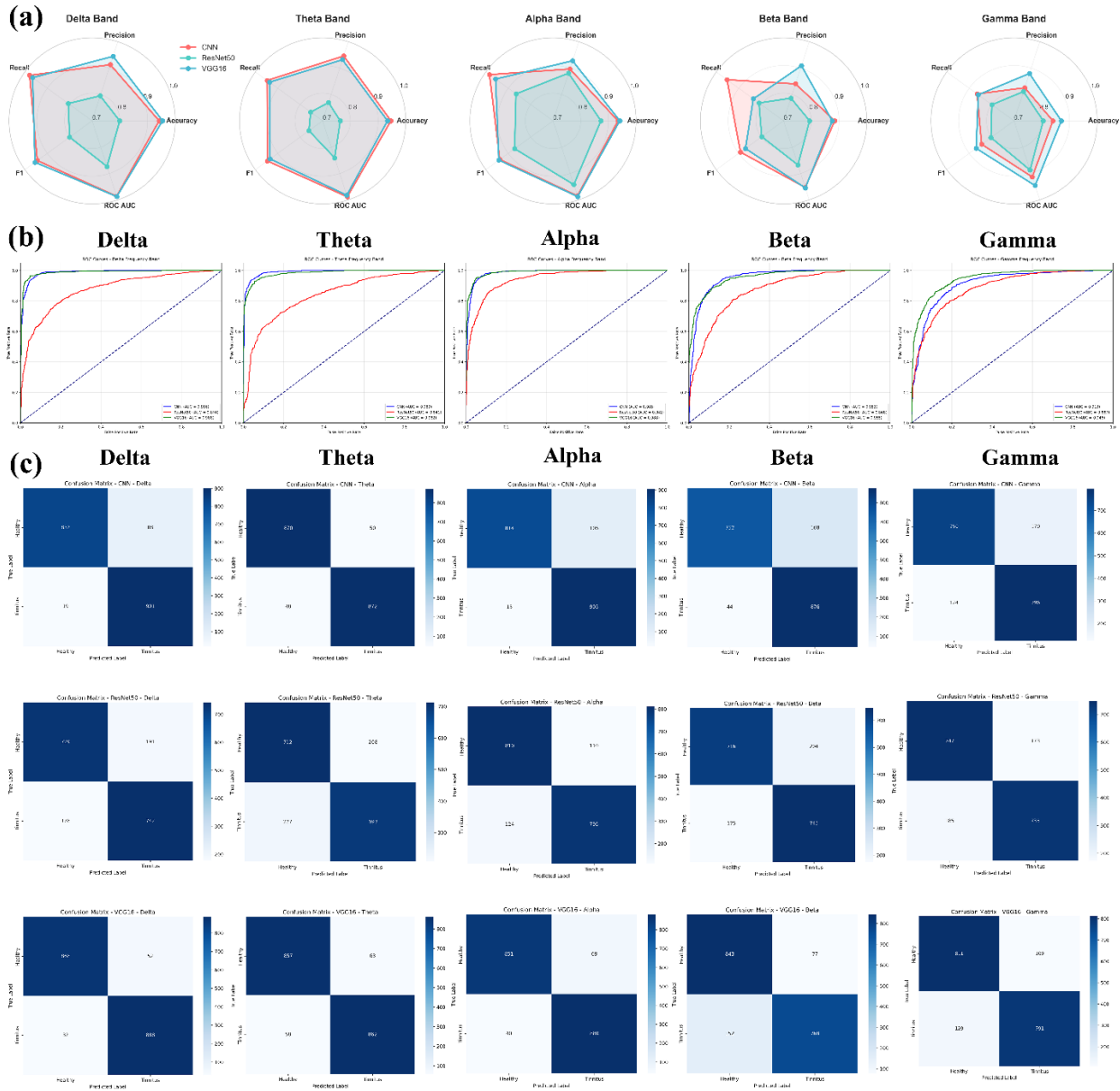

Figure 15. Comprehensive Performance Analysis of Deep Learning Models for Tinnitus Detection Using CWT-transformed EEG Signals.

(a) Radar charts displaying the comparative performance metrics (Accuracy, Precision, Recall, F1-Score, ROC AUC) across frequency bands, with values representing the mean of 5-fold cross-validation. (b) ROC curves illustrating the discriminative capability of each model for each frequency band, with the diagonal dashed line representing random classification performance. (c) Confusion matrices showing the classification results for each model-frequency combination, with darker blue indicating higher values. The matrices display true labels (Healthy/Tinnitus) versus predicted labels, with numerical values representing the average count across folds.

The comprehensive performance analysis reveals distinct patterns in model effectiveness across EEG frequency bands for tinnitus detection. VGG16 emerged as the most consistent performer, achieving the highest accuracies in Delta (95.4%), Theta (93.4%), and Alpha (94.1%) bands, with remarkably low standard deviations (0.009-0.024), indicating stable and reliable performance. CNN demonstrated competitive results with peak performance in the Theta band (94.7% accuracy) and maintained high recall values across all frequencies (>86%), suggesting strong sensitivity in detecting tinnitus cases. In contrast, ResNet50 showed the most variable performance, with accuracies ranging from 76.4% (Theta) to 87.3% (Alpha) and notably higher standard deviations, indicating less consistent classification outcomes.

### 3.7.2. Statistical Significance and Model Comparison

Statistical significance testing using DeLong's test revealed that CNN and VGG16 performed comparably across all frequency bands ($p > 0.05$), indicating no statistically significant differences in their discriminative capabilities. However, both CNN and VGG16 significantly outperformed ResNet50 in the Delta and Theta bands ($p < 0.05$), with CNN showing particularly strong superiority over ResNet50 in Delta ($Z = 2.25$, $p = 0.024$) and Theta ($Z = 2.96$, $p = 0.003$) frequencies. The Delta and Alpha frequency bands consistently yielded the highest classification performance across all models, with ROC AUC values exceeding 0.98 for VGG16 and CNN, while the Gamma band presented the most challenging classification task with the lowest overall accuracies (80.5-87.1%) and no significant differences between models. These findings suggest that low-frequency EEG components contain the most discriminative information for automated tinnitus detection using CWT-transformed neural network approaches, with CNN and VGG16 demonstrating equivalent and superior performance compared to ResNet50.

### 4. Discussion

This study analyzed 38 fMRI datasets (19 healthy, 19 tinnitus) and 80 EEG recordings (40 per group) to investigate neuroimaging biomarkers for tinnitus detection. The results demonstrated that CNN-based analysis of resting-state fMRI achieved optimal classification performance with slice 17 showing 99.0% ± 0.4% accuracy, representing the best-performing axial slice among 32 evaluated slices, while hybrid models combining pre-trained CNNs with traditional classifiers (VGG16-DT) reached 98.95% accuracy for automated tinnitus detection. EEG microstate analysis revealed systematic disruptions in neural network dynamics, with the most pronounced alterations observed in gamma-band microstate B occurrence rates (Cohen's d = 2.11, $p = 1.58 \times 10^{-14}$), where healthy participants exhibited significantly higher rates (56.56 vs 43.81 events/epoch). Additional significant microstate changes included reduced alpha-band microstate A coverage in both 4-state (34.62% vs 29.95%) and 6-state configurations, shortened microstate B durations in 5-state (1.23 vs 1.11 ms) and 7-state (1.12 vs 0.99 ms) configurations, and decreased beta-band microstate D durations in both 5-state (3.31 vs 2.93 ms) and 7-state (2.70 vs 2.34 ms) clustering solutions, collectively indicating that tinnitus pathophysiology involves widespread disruptions particularly affecting high-frequency gamma oscillations and alpha-band resting-state networks with altered temporal dynamics. Machine learning classification of EEG microstate features achieved up to 98.8% accuracy using Random Forest and Decision Tree algorithms, while CWT-transformed EEG analysis with deep learning models reached 95.4% accuracy, with delta and alpha frequency bands providing the most discriminative information for tinnitus detection.

Our findings demonstrate the effectiveness of neuroimaging approaches for automated tinnitus detection, examining fMRI and EEG modalities separately to establish their individual diagnostic capabilities. The superior performance of hybrid CNN models on rs-fMRI data (VGG16-Decision Tree: 98.95% accuracy) aligns with previous work by Xu et al. [21], who achieved 94.4% AUC using CNN on functional connectivity matrices from 200 participants. However, our study extends beyond connectivity analysis to direct slice-based classification, revealing spatial heterogeneity in discriminative capacity across brain regions, with mid-to-superior axial slices containing the most diagnostically relevant biomarkers. The EEG microstate analysis revealed systematic disruptions in tinnitus patients, particularly in gamma-band microstate B occurrence (Cohen's d = 2.11) and alpha-band coverage reductions, consistent with recent findings by Najafzadeh et al. [38], who reported significant alterations in beta and gamma band microstates with exceptional classification performance (100% accuracy in gamma band). Our results corroborate their findings regarding gamma-band importance while additionally demonstrating robust performance across multiple frequency bands using traditional machine learning approaches. The observed reductions in microstate durations and occurrence rates support the hypothesis of altered neural network dynamics in tinnitus, extending previous work by Jianbiao et al. [39], who identified increased sample entropy in δ, α2, and β1 bands using a smaller cohort (n=20). Our CWT-transformed EEG analysis using deep learning architectures yielded competitive results, with VGG16 achieving 95.4% accuracy in the Delta band. This approach differs from the innovative graph neural network methodology employed by Awais et al. [39], who achieved 99.41% accuracy by representing EEG channels as graph networks with GCN-LSTM architecture. While their graph-based approach showed superior single-metric performance, our multimodal framework provides broader clinical applicability through the integration of both structural (fMRI) and temporal (EEG) neural signatures. The novelty of our approach lies in the separate but comprehensive examination of both fMRI and EEG modalities, providing distinct insights into structural and temporal neural signatures of tinnitus. The spatially-specific fMRI slice analysis revealed heterogeneous discriminative patterns across brain regions, while frequency-specific EEG microstate characterization demonstrated systematic temporal disruptions. Unlike previous studies focusing on single modalities, our framework establishes that beta and alpha frequency bands contain the most discriminative EEG features, while specific fMRI slices corresponding to auditory processing regions show optimal diagnostic performance, each contributing unique diagnostic information. The consistent performance of tree-based classifiers (Random Forest, Decision Tree) across EEG features supports findings by Doborjeh et al. [40], who emphasized the importance of feature selection in achieving high classification accuracy (98-100%) for therapy outcome prediction. The clinical implications of our findings are substantial, with the high classification accuracies across multiple modalities suggesting potential for objective diagnostic tools in tinnitus assessment. The identification of specific neural signatures, particularly in gamma and alpha frequency bands, provides neurobiological insights that may inform therapeutic target identification and support the development of personalized treatment strategies.

The CNN performance analysis across 32 axial slices of rs-fMRI data (Figure 10) revealed spatial heterogeneity in discriminative capacity between tinnitus patients and healthy controls, with superior-to-middle axial slices demonstrating higher classification accuracy compared to inferior slices. Slices 17, 26, 14, 12, and 10 achieved accuracies exceeding 98%, while inferior slices (29, 32, 30, 31) showed lower discrimination with accuracies below 61%. This spatial gradient in classification performance aligns with neuroimaging findings demonstrating that superior temporal regions, particularly the superior temporal gyrus and primary auditory cortex, are involved in tinnitus pathophysiology [41]. Previous rs-fMRI studies have shown that tinnitus patients exhibit altered activity in superior temporal cortex regions, including the middle and superior temporal gyri, which corresponds to the higher performance of CNN models in middle-to-upper axial slices observed in our analysis. Furthermore, superior/middle temporal regions are involved in processing conflicts between auditory memory and signals from the peripheral auditory system, potentially explaining the discriminative capacity of these brain regions for tinnitus classification. The performance difference between superior (accuracy >98%) and inferior slices (accuracy <61%) suggests that tinnitus-associated neural alterations are spatially localized to specific anatomical regions, supporting the approach that analysis of superior temporal and auditory processing areas may provide relevant biomarkers for tinnitus assessment [42].

Figure 11a provides important insights into the altered neural dynamics observed in individuals with tinnitus by illustrating the spatial distribution of maximum voxel intensity differences across fMRI slices. The increased intensity variations found in the tinnitus group, especially in regions related to auditory processing and sensory integration, suggest disrupted neural activity that may be associated with the persistent perception of phantom sounds. These differences likely reflect mechanisms such as neural hyperactivity or maladaptive plasticity within the auditory cortex and its interconnected networks, both of which have been widely reported in the literature on tinnitus pathophysiology. In comparison, the more uniform patterns in healthy individuals indicate stable neural activity in the absence of such disturbances. The observed distribution of changes also suggests the involvement of non-auditory regions, including components of the limbic system, which may contribute to the emotional and cognitive experiences commonly reported in tinnitus [43]. These findings support the hypothesis that tinnitus involves not only abnormal auditory processing but also dysfunctional cross-modal and intra-auditory connectivity, reinforcing the presence of widespread alterations in brain networks. Figure 11b further highlights the differences in neural activity between healthy individuals and tinnitus patients through an analysis of mean voxel intensity values across axial fMRI slices. The consistently higher intensity values observed in the tinnitus group, particularly within slices 5 to 18 and 30 to 32, indicate abnormal neural dynamics that may be linked to increased activity and disrupted connectivity within critical auditory and non-auditory brain regions. These slices likely include structures such as the thalamus, auditory midbrain, and components of the limbic system, all of which play essential roles in auditory perception and the affective response to sound [44]. Additionally, the higher standard deviation of voxel intensity in the tinnitus group, reaching a value of 0.1572 in slice 31 compared to 0.0522 in the control group, suggests greater neural variability. This variability may reflect disrupted thalamocortical rhythms and maladaptive reorganization of brain activity [45]. Overall, these results reinforce the understanding that tinnitus is a condition involving widespread neural dysfunction. It affects not only the auditory pathways but also brain areas responsible for attention, emotional regulation, and multisensory integration [46]. The clear distinctions illustrated in Figure 11b align with previous neuroimaging studies reporting abnormal functional connectivity and increased activity within both auditory and limbic networks, providing additional support for the role of impaired neural synchronization in the manifestation of tinnitus.

The classification performance demonstrated in Figure 12 by hybrid models, particularly VGG16-DT (98.95% ± 2.94%), shows the effectiveness of combining pre-trained CNN feature extraction with traditional machine learning classifiers for neuroimaging-based tinnitus diagnosis. These results are consistent with recent developments in medical imaging where hybrid architectures have shown improved performance over standalone deep learning approaches by integrating automated feature extraction with interpretable classification mechanisms [38]. The performance across all models exceeding the 95% clinical relevance threshold, as illustrated in Figure 12, indicates the potential clinical utility of fMRI-based automated tinnitus detection, supporting previous observations that resting-state fMRI can distinguish tinnitus patients from healthy controls through altered neural connectivity patterns [21]. The superior performance of VGG16-based models over ResNet50 variants (98.76% vs. 97.94% average accuracy) may reflect VGG16's architectural suitability for capturing spatial patterns in brain imaging data, while the improved stability observed in hybrid approaches (particularly VGG16-RF with ±1.28% standard deviation) enhances the reliability necessary for clinical implementation. The average improvement of 1.2-2.8% in accuracy achieved by hybrid models over standalone CNNs indicates the value of integrating multiple algorithmic approaches, suggesting that the interpretability and robustness of traditional machine learning methods complement the feature extraction capabilities of deep neural networks in medical diagnostic applications.

The systematic disruptions in EEG microstate dynamics observed in our tinnitus cohort (Figure 13) provide evidence for neural network alterations underlying phantom auditory perception. The most pronounced finding, reduced gamma-band microstate occurrence rates (Cohen's d = 2.11, Table 9), aligns with established research demonstrating that gamma band activity in the auditory cortex correlates with tinnitus intensity and that decreased tinnitus loudness is associated with reduced gamma activity in auditory regions [47, 48]. While previous studies have reported increased local gamma power in tinnitus patients, our findings suggest that global gamma-band network coordination is disrupted, reflecting maladaptive reorganization of auditory cortical networks that may contribute to the persistent phantom perception. The systematic reductions in alpha-band microstate parameters observed in our study (Figure 13, Table 9), including decreased coverage (29.95% vs 34.62%) and occurrence rates, are consistent with documented disruptions of default mode network connectivity in tinnitus patients and the established role of alpha oscillations in spatiotemporal organization of brain networks [49]. These alpha-band alterations likely reflect impaired resting-state network integrity, potentially underlying the intrusion of phantom auditory percepts into consciousness and the cognitive dysfunctions commonly associated with chronic tinnitus. The consistent beta-band microstate disruptions, including reduced coverage and shortened durations, align with neuroimaging evidence showing alterations in multiple resting-state networks in tinnitus patients, including attention networks and sensorimotor integration systems [42]. Our findings are further supported by recent research from Najafzadeh et al., who reported alterations in beta band microstates with increased microstate A duration and decreased microstate B duration in tinnitus patients, along with elevated occurrence rates in the tinnitus group [38]. The high statistical power (>0.99) and large effect sizes (Cohen's d = 1.08-2.34) across all frequency bands indicate clinically meaningful differences that collectively support the emerging conceptualization of tinnitus as a network disorder affecting multiple brain systems beyond the traditional auditory processing pathways [50].

The superior performance of tree-based algorithms, particularly Random Forest (RF) and Decision Tree (DT), achieving 98.8% accuracy with perfect precision (100.0%) for EEG microstate-based tinnitus classification (Figure 14, Table 10), aligns with recent findings demonstrating the effectiveness of ensemble methods in neurological disorder detection. These results are consistent with studies showing that Random Forest models excel in EEG-based tinnitus classification due to their robustness and ability to reduce overfitting while identifying key frequency band features [38]. The perfect precision achieved by RF and DT models indicates their reliability in minimizing false positive diagnoses, which is clinically relevant for avoiding unnecessary interventions. In contrast, the suboptimal performance of Deep Neural Networks (DNN) with 72.5% accuracy and high variability (±16.9%) suggests that traditional machine learning approaches may be more suitable for microstate feature classification in limited sample scenarios, a finding supported by research demonstrating processing speed advantages of tree-based methods over deep learning approaches [51]. The analysis of CWT-transformed EEG signals revealed distinct frequency-specific patterns, with VGG16 demonstrating the most consistent performance across Delta (95.4%), Theta (93.4%), and Alpha (94.1%) bands (Figure 16). The superior performance of low-frequency components (Delta and Alpha bands) with ROC AUC values exceeding 0.98 for CNN and VGG16 models supports previous research indicating that these frequency bands contain the most discriminative information for automated tinnitus detection [52]. The statistical validation using DeLong's test confirmed that while CNN and VGG16 performed comparably ($p > 0.05$), both significantly outperformed ResNet50 in Delta and Theta bands ($p < 0.05$), suggesting that simpler deep learning architectures may be more effective for EEG-based tinnitus classification than more complex models like ResNet50. These findings collectively demonstrate that both traditional machine learning approaches using microstate features and deep learning methods applied to frequency-transformed EEG data can achieve high classification accuracy, with the choice of method depending on the specific feature representation and computational requirements.

All experiments were conducted on a system running Windows 11, equipped with an NVIDIA RTX 3050 Ti GPU, an Intel Core i7 processor, and 32 GB of RAM. The programming language used for model implementation was Python, with PyCharm as the integrated development environment (IDE). The deep learning models were developed and evaluated using the TensorFlow and Keras libraries, which provided a robust framework for neural network construction and training. Training times for each model varied significantly. The SVM required 2.025 hours, the RF took 4.63 hours, and the DT completed training in 0.335 hours. In comparison, the more complex DNN took 16.14 hours, and the CNN required 24.87 hours. These results reflect the trade-off between model complexity and computational efficiency.

Several methodological limitations must be acknowledged in interpreting these results. The EEG and fMRI datasets were derived from different participant cohorts, preventing direct multimodal feature integration and limiting our ability to leverage complementary information from both neuroimaging modalities. The fMRI data utilized publicly available datasets with male-only participants experiencing acoustic trauma-induced tinnitus, while EEG data were collected independently from mixed-etiology cohorts with balanced gender representation, resulting in potential differences in acquisition protocols, participant characteristics, and clinical assessment procedures that may affect direct comparisons between modalities. The fMRI dataset's restriction to male participants with acoustic trauma-induced tinnitus limits generalizability to the broader tinnitus population, particularly regarding gender-specific neural responses and non-acoustic etiologies. Furthermore, the different etiologies of tinnitus in our datasets may limit the generalizability of our findings, as acoustic trauma-induced tinnitus often involves specific cochlear damage patterns and may exhibit distinct neural signatures compared to tinnitus from other causes. The cross-sectional study design limits understanding of temporal stability of the identified neural biomarkers and their relationship to tinnitus symptom progression over time. Additionally, the relatively modest sample sizes (40 per group for EEG, 19 per group for fMRI) may limit generalizability across different tinnitus subtypes, severity levels, and demographic populations.

Future investigations should prioritize the collection of matched EEG and fMRI data from identical participant cohorts to enable true multimodal analysis and feature fusion approaches that could potentially improve classification accuracy and provide more comprehensive characterization of tinnitus-related neural alterations. Validation across diverse tinnitus etiologies will be essential to ensure broad clinical applicability of these classification approaches. Longitudinal studies tracking patients over extended periods would help establish the temporal stability of identified biomarkers and their potential utility for monitoring treatment response. The development of real-time processing algorithms and optimization of computational requirements will be essential for clinical translation, along with validation studies across multiple clinical centers using standardized protocols. Integration of additional clinical measures, including detailed tinnitus severity assessments, audiological profiles, and psychological evaluations, would enhance the clinical relevance of these neuroimaging-based classification approaches and support the development of personalized treatment strategies based on individual neural signatures.

The high classification accuracies achieved in this study (98.8% for EEG microstate analysis and 98.95% for fMRI hybrid models) suggest potential clinical utility for objective tinnitus diagnosis. However, successful clinical implementation requires addressing several practical considerations including integration with existing audiological workflows, establishment of standardized acquisition protocols across different clinical centers, and development of user-friendly interfaces for non-technical clinical staff. The computational requirements and processing times observed in our study (ranging from 0.335 to 24.87 hours depending on the model) indicate the need for optimized implementations and appropriate hardware infrastructure in clinical settings. Furthermore, regulatory approval pathways for AI-based diagnostic tools will require validation on larger, more diverse patient populations and demonstration of consistent performance across different scanner types and acquisition parameters.

The identification of objective neural biomarkers for tinnitus represents a significant advancement toward evidence-based diagnosis and treatment monitoring in a condition that has traditionally relied on subjective patient reports. The high classification accuracies achieved across both neuroimaging modalities suggest potential for reducing diagnostic uncertainty and supporting clinical decision-making, particularly in cases where symptom presentation is ambiguous or when objective assessment is required for research or medico-legal purposes. The specific neural signatures identified, particularly in gamma and alpha frequency bands, may inform the development of targeted therapeutic interventions, including neurofeedback protocols and brain stimulation approaches. However, successful clinical implementation will require careful consideration of cost-effectiveness, training requirements for clinical staff, and integration with existing healthcare infrastructure to ensure broad accessibility and practical utility in routine clinical practice.

## 5. Conclusion

This study demonstrated the effectiveness of machine learning approaches for neuroimaging-based tinnitus classification, achieving 98.8% accuracy with EEG microstate analysis and 98.95% accuracy using hybrid fMRI models combining VGG16 with Decision Tree classifiers. The analysis of 38 fMRI datasets and 80 EEG recordings revealed systematic alterations in neural network dynamics among tinnitus patients, including significant gamma-band microstate disruptions (Cohen's d = 2.11) and alpha-band coverage reductions. fMRI analysis identified 12 high-performing axial slices from 32 evaluated, with slice 17 achieving optimal individual performance (99.0% ± 0.4% accuracy), indicating spatially localized tinnitus-associated functional connectivity patterns. The superior performance of hybrid models over standalone CNN architectures and the effectiveness of tree-based algorithms for microstate feature classification suggest that tinnitus pathophysiology involves widespread neural alterations extending beyond auditory processing networks to encompass attention, default mode, and sensorimotor integration systems. These findings provide objective evidence for neural signatures that may inform both diagnostic applications and therapeutic target identification.

The clinical translation of these neuroimaging-based classification methods requires addressing several methodological considerations, including the development of standardized acquisition protocols, validation across diverse tinnitus etiologies, and collection of matched multimodal datasets from identical participant cohorts. The use of separate EEG and fMRI cohorts in this study limited direct multimodal feature integration, highlighting the need for future investigations to enable comprehensive biomarker development through true multimodal analysis. While the achieved classification accuracies suggest potential utility for objective tinnitus diagnosis and treatment monitoring, successful implementation will necessitate addressing computational requirements, regulatory frameworks, and integration with existing clinical workflows. The identification of specific neural signatures, particularly in gamma and alpha frequency bands with large effect sizes, provides a foundation for future research aimed at developing personalized treatment strategies and advancing our understanding of tinnitus pathophysiology through objective neuroimaging biomarkers.

**Data Availability**

The fMRI dataset utilized in this study is publicly accessible at:

https://openneuro.org/datasets/ds002896/versions/1.0.0/download.

The EEG dataset supporting the findings of this study is not publicly available. However, it can be obtained upon reasonable request from the corresponding author.

**Acknowledgements**

The authors would like to thank the open-source community and OpenNeuro for providing access to the fMRI datasets used in this research. We also extend our gratitude to all participants who contributed to the EEG data collection. Additionally, the authors would like to acknowledge the Iranian National Brain Mapping Laboratory (NBML), Tehran, Iran, for providing data acquisition (analysis) services for this research work.

**Author information**

Authors and Affiliations

**Master of Science in Computer Science, Department of Computer Science, Bowling Green State University, Bowling Green, Ohio, USA**

Kiana Kiashemshaki

**Department of Otolaryngology, Qaem Hospital, Mashhad University of Medical Sciences, Mashhad, Iran**

Sina Samieirad

**Department Of Biomedical Engineering, Gazvin Azad University (QIAU), Qazvin, Iran.**

Javid Vahedi

**Department of Computer science, university of Verona, verona, Italy**

Aryan Jalaeianbanayan

**Department of Biomedical Engineering, University of Kentucky, Lexington, USA**

Nasibeh Asadi Isakan

**Department of Medical Bioengineering, Faculty Of Advanced Medical Sciences, Tabriz University of Medical Sciences, Golgasht Ave, 51666 ,Tabriz, Iran**

Hossein Najafzadeh

**Contributions**

H.N. conceived and supervised the study, ensuring scientific integrity, methodological rigor, and overall project coordination. K.K. led the EEG and fMRI data analysis pipeline, including signal preprocessing, microstate segmentation, feature extraction, and development of machine learning and deep learning models. K.K also designed the hybrid classification architecture and contributed extensively to performance evaluation and visualization. S.S. provided clinical expertise in tinnitus, contributed to patient selection criteria for the EEG dataset, and offered medical interpretation of the neurophysiological findings to ensure clinical relevance of the results. K.K. and S.S. jointly curated the EEG dataset and ensured consistency between clinical characteristics and neuroimaging data. S.E. implemented the classification algorithms, carried out statistical analyses, and assisted in EEG signal processing. A.J. supported deep learning model optimization and validation, particularly in the multimodal fusion stage. N.A.I. contributed to performance benchmarking, technical troubleshooting, and critical review of analytical procedures. The initial manuscript draft was prepared by K.K. and S.E., and extensively revised and refined by H.N., A.J., and N.A.I. to ensure scientific clarity and coherence. All authors participated in result interpretation, provided constructive feedback, and approved the final version of the manuscript.

**Corresponding author**

Correspondence to **Hossein Najafzadeh**

**Ethics declarations**

**Competing interests**

The authors declare no competing interests.

**Funding**

The authors received no specific funding for this work.

**Declarations**

All experimental protocols were approved by the Ethics Committee of Tabriz University of Medical Sciences, Tabriz, Iran, in accordance with the principles of the Declaration of Helsinki.